\documentclass[10pt] {article}

\AtBeginDocument{%
  }

\usepackage{balance}

\usepackage[usenames,dvipsnames,svgnames,x11names]{xcolor}
\usepackage{color}
\definecolor{gray}{rgb}{0.4,0.4,0.4}
\definecolor{lightgray}{rgb}{0.9,0.9,0.9}
\definecolor{darkblue}{rgb}{0.0,0.0,0.6}
\definecolor{maroon}{rgb}{0.5,0,0}
\definecolor{darkgreen}{rgb}{0,0.6,0}
\usepackage{bbding}

\usepackage[T1]{fontenc}
\usepackage[normalem]{ulem}

\usepackage[export]{adjustbox}

\newcommand{\bluecol}[1]{\textcolor{darkblue}{#1}}

\newcommand{\claim}[1]{\vspace{-0.08in}\subsubsection*{\textnormal{\textit{\bluecol{#1}}}}}
\usepackage{listings}

\lstdefinelanguage{XML}
{
basicstyle=\ttfamily\footnotesize,
  morestring=[b]",
  moredelim=[s][\bfseries\color{Maroon}]{<}{\ },
  moredelim=[s][\bfseries\color{Maroon}]{</}{>},
  moredelim=[l][\bfseries\color{Maroon}]{/>},
  moredelim=[l][\bfseries\color{Maroon}]{>},
  morecomment=[s]{<?}{?>},
  morecomment=[s]{<!--}{-->},
  commentstyle=\color{gray},
  stringstyle=\color{blue},
  identifierstyle=\color{red}
}

\usepackage{moreverb}

\usepackage[nounderscore]{syntax}

\graphicspath{{./figures/}}
\DeclareGraphicsExtensions{.pdf}
\usepackage[export]{adjustbox}

\usepackage[cmex10]{amsmath}
\usepackage{amssymb}
\usepackage{mathtools}
\usepackage{amsthm}
\usepackage{amsfonts}

\DeclareMathOperator*{\subto}{s.t.}
\DeclareMathOperator*{\sel}{select}

\usepackage{subfig}
\usepackage{algorithmicx}
\usepackage{algpseudocode}
\usepackage[ruled]{algorithm}
\usepackage{setspace}
\definecolor{light-gray}{gray}{0.75}
\algrenewcommand{\algorithmiccomment}[1]{\hskip3em{{\footnotesize \textcolor{light-gray}{$\blacktriangleright$}}} #1}

\usepackage{multirow}
\usepackage{rotating}
\usepackage{booktabs}
\usepackage{colortbl} 
\usepackage{tablefootnote}
\usepackage{array}
\newcolumntype{L}[1]{>{\raggedright\let\newline\\\arraybackslash\hspace{0pt}}m{#1}}
\newcolumntype{C}[1]{>{\centering\let\newline\\\arraybackslash\hspace{0pt}}m{#1}}
\newcolumntype{R}[1]{>{\raggedleft\let\newline\\\arraybackslash\hspace{0pt}}m{#1}}

\usepackage[colorlinks,bookmarksopen,bookmarksnumbered,citecolor=red,urlcolor=red]{hyperref}

\usepackage{xspace}

\usepackage{enumitem}

\usepackage{bbding}
\usepackage{pifont}
\usepackage{wasysym}
\usepackage{oplotsymbl}

\usepackage{blindtext}

\newcommand{\mobilenet}{MobileNet\xspace}
\newcommand{\resnet}{ResNet\xspace}
\newcommand{\yolo}{YOLO\xspace}
\newcommand{\lstm}{LSTM\xspace}
\newcommand{\bert}{BERT\xspace}
\newcommand{\papertitle}{Fulcrum\xspace}

\newcommand*\circled[1]{\tikz[baseline=(char.base)]{
            \node[shape=circle,draw,inner sep=1pt] (char) {#1};}}

\begin{document}
\title{\textit{\papertitle:} Optimizing Concurrent DNN Training and Inferencing on Edge Accelerators}

\author{Prashanthi S. K., Saisamarth Taluri, Pranav Gupta, \\ Amartya Ranjan Saikia, Kunal Kumar Sahoo, Atharva Vinay Joshi,\\
Lakshya Karwa, Kedar Dhule and Yogesh Simmhan\\
Department of Computational and Data Sciences,\\
Indian Institute of Science, Bangalore 560012 India\\
Email: \{prashanthis, simmhan\} @iisc.ac.in}

\date{}

\maketitle

\begin{abstract}
The proliferation of GPU accelerated edge devices like Nvidia Jetsons and the rise in privacy concerns are placing an emphasis on concurrent DNN training and inferencing on edge devices. Inference and training have different computing and QoS goals. But edge accelerators like Jetson do not support native GPU sharing and expose 1000s of power modes. This requires careful time-sharing of concurrent workloads to meet power--performance goals, while limiting costly profiling.
In this paper, we design an intelligent time-slicing approach for concurrent DNN training and inferencing on Jetsons. We formulate an optimization problem to interleave training and inferencing minibatches, and decide the device power mode and inference minibatch size, while maximizing the training throughput and staying within latency and power budgets, with modest profiling costs. We propose \textit{GMD}, an efficient multi-dimensional gradient descent search  which profiles just $15$ power modes; and \textit{ALS}, an Active Learning technique which identifies reusable Pareto-optimal power modes, but profiles $50$--$150$ power modes. We evaluate these within our \textit{\papertitle} scheduler for $273,000+$ configurations across $15$ DNN workloads. We also evaluate our strategies on dynamic arrival inference and concurrent inferences. ALS and GMD outperform simpler and more complex baselines with larger-scale profiling. Their solutions satisfy the latency and power budget for $>97\%$ of our runs, and on average are within $7\%$ of the optimal throughput. 
\end{abstract}

\section{Introduction}

Edge devices are being deployed alongside sensors in smart environments like digital health~\cite{health_edge}, smart cities~\cite{IoT_edge} and autonomous vehicles~\cite{AV_edge}, close to the data source, typically to assist with data acquisition and on-device inferencing.
Recent advances have seen GPUs and TPUs hosted in these edge devices while still retaining a low power envelope and small form factor~\cite{prashanthi2023sigmetrics,matei_europar}.
While accelerators like Google Coral and Intel Movidius are designed for light-weight Deep Neural Network (DNN) inference tasks~\cite{Edge}, Nvidia's Jetson edges have GPUs that approach workstation capability. E.g., the Jetson AGX Orin has a 12-core ARM CPU, an Nvidia Ampere GPU with 2048 CUDA cores, and 32GB of RAM shared between CPU and GPU, which is comparable to an RTX 3080 Ti workstation. 
As a result, such edge accelerators can go beyond just inferencing to also perform on-device training over the local data. This can preserve privacy using \textit{federated learning}~\cite{pmlr_mcmahan_FL} and also reduce data movement to the cloud, e.g., during \textit{continuous learning}~\cite{shmelkov2017incremental,ekya-stoica-nsdi} over evolving data.

\paragraph{Challenges}
On-device edge Artificial Intelligence (AI) can include DNN training or inferencing, separately~\cite{edge-comp-survey} or concurrently~\cite{ekya-stoica-nsdi, continual_sensys23}. 

\textit{Inference workloads} are latency-sensitive and lightweight, as they usually execute only the forward pass of DNN training. They may run continuously or interactively, responding to the user or environment. \textit{Training workloads} are compute-heavy and throughput sensitive. They may run periodically, be triggered upon data drift, or launched for model-tuning. 
E.g., safety camera networks in a smart city with co-located edge devices can perform vehicle-count inferencing continuously to inform traffic signaling but retain models daily in a federated manner~\cite{federated_traffic}.
The actual inferencing and training models that are run may also change as newer models emerge. 

So, there is a critical need to efficiently execute such concurrent, heterogeneous and evolving AI inferencing and training workloads on edge accelerators.

Contemporary edge accelerators like Nvidia Jetson do not support native GPU sharing using CUDA \textit{Multi Process Service (MPS)}~\cite{MPS} or \textit{Multi-Instance GPU (MIG)}~\cite{MIG}, present in server-grade GPUs. So, when we simultaneously run training and inferencing on Jetsons, they 

\textit{time-share} the GPU if run from separate CPU processes~\cite{edge_gpustreams} and \textit{space-share} the GPU if run using multiple CUDA streams from a single CPU process~\cite{edge_gpustreams, tx2scheduling_rtss, kalmia_infocom}. When time-shared, the NVIDIA GPU scheduler interleaves the executions of kernels, with time-slices allocated at $\mu\text{s}$ granularity. This native time-sharing can be sub-optimal, with variable inference latencies.

E.g., in \textit{Cfg $8$} of Figure~\ref{fig:intconc}, $65\%ile$ of inference requests violate the budget and the training throughput is $10.8\%$ below a nominal optimal. When training and inference are space-shared using two different CUDA streams, training throughput improves but inference latency still has high variability due to non-deterministic resource blocking delays (Figure~\ref{fig:intconc}). This happens even if inference is assigned a higher-priority CUDA stream~\cite{multiinf_taas23, pantheon_mobisys24}. Such \textit{latency variability and violations} for inference may be unacceptable for mission critical domains such as biomedical devices in smart healthcare~\cite{healthcare_dnn} and intelligent controllers in Industry 4.0~\cite{iiot_dnninf}.

Further, edge devices can be deployed in field settings such as autonomous vehicles~\cite{drone}, smart cities~\cite{IoT_edge} and smart agriculture~\cite{sajal-smart-agri-edge}. These may impose \textit{constraints on power}, e.g., to avoid overheating of an IP-67 packaged device deployed to count vehicles,
or \textit{energy}, e.g., to avoid draining the battery on-board a drone. \textit{So, it is crucial to identify an optimal configuration for the edge device and the concurrent DNN workloads running on them to meet the user's inference latency, training throughput and power goals.}

Nvidia Jetsons expose 1000s of \textit{power mode configurations}, which control the active CPU cores, and the CPU, GPU and memory frequencies at fine granularity. These offer a meaningful control knob to help meet the Quality of Service (QoS) goals. However, profiling each power mode is time-consuming, taking 10s of seconds; this also needs to be done for each new DNN model and device.  Automatically selecting the best power mode to meet the application's needs is an open problem. Similarly, minibatch size of inferencing  can affect the inferencing latency and, as we show, also impact the training throughput for concurrent workloads.

\paragraph{Gaps} There has been much work on optimizing the power and performance for training on GPU servers~\cite{gandiva_osdi, antman_osdi20, MURI_sigcomm, zeus_nsdi23, gslice_socc20}. Some use GPU partitioning such as MPS and MIG~\cite{gslice_socc20}, and others use knobs such as the GPU power limit~\cite{zeus_nsdi23}. These knobs are not available on edge GPUs. None of them simultaneously control the CPU, GPU and memory frequencies, which together affect the power and performance of edge accelerators~\cite{prashanthi2023sigmetrics}. Therefore, optimization studies on the server are not directly applicable.

There is literature on characterizing, predicting and optimizing AI inference workloads on edge devices~\cite{matei_europar, mirage, alert_atc20}. In contrast, training studies on edge devices are limited~\cite{PowerTrain}. There are a few works on concurrent training and inference on servers~\cite{ekya-stoica-nsdi,lyra_eurosys23}, but they too rely on GPU sharing. \textit{To the best of our knowledge, there is no work on optimizing concurrent training and inference on edge accelerators.} 

\paragraph{Contributions} In this article, we address this gap by intelligently controlling the power modes, inferencing minibatch sizes, and scheduling of tasks to optimize the performance of evolving training and inference workloads, executed independently and concurrently, to meet the QoS goals on inference latency, training throughput and power budget. We make the following specific contributions:
\vspace{-0.05in}
\begin{enumerate}[leftmargin=*]%
\item We discuss multiple existing approaches to address this problem on NVIDIA Jetson edge accelerators, and find our proposed \textit{managed interleaved execution} of mixed workloads as a viable strategy 
(\S~\ref{sec:approach}). 
\item We formulate a \textit{multi-criteria optimization problem} to tune the power mode and inference minibatch size ($bs$)~\footnote{While training is also done in minibatches, its minibatch size is a user-defined hyper-parameter that affects training accuracy. We do not change it.} to meet the user-defined power and inference latency constraints while maximizing the training throughput, for standalone and concurrent executions (\S~\ref{sec:problem}). Secondary goals are to lower the inference latency and profiling costs.
\item We develop two novel approaches to solve this problem: a \textit{Gradient-descent based Multi-Dimensional search (GMD) strategy} 
that profiles just a few power modes to solve a specific problem configuration; and an \textit{Active-Learning based Sampling (ALS) strategy} that uses greedy sampling to profile the Pareto optimal power modes and satisfies a wide variety of problem configurations (\S~\ref{sec:strategies}).

We define the \textit{\papertitle} scheduler to implement managed interleaving.

\item We perform a \textit{rigorous evaluation} of GMD and ALS using \papertitle for 
$273k+$ workload configurations, spanning $\approx 40$ power budgets, $\approx 95$ latency budgets, $\approx 10$ inference arrival rates and $15$ DNN workloads ($5$ training, $5$ inferencing and $5$ concurrent training and inference pairs) on an NVIDIA Jetson Orin AGX (\S~\ref{sec:setup},~\ref{sec:results}). We also validate our solutions on $2$ pairs of concurrent inference workloads, and $3$ different inference workloads with dynamic arrival rates. We compare against three simple and complex baselines.
GMD and ALS find solutions over $97\%$ of the time, always meet latency and power budgets, and are within $7\%$ of the nominal optimal solution, on average.
\end{enumerate} 
\vspace{-0.05in}
Besides these, we further motivate the problem in \S\ref{sec:motivation}, discuss related work in \S~\ref{sec:related}, and offer our conclusions in \S~\ref{sec:conclude}.

This article builds upon our prior 2-page poster paper~\cite{ccgridw} that investigated the possibility of concurrent execution of training and inference, and characterized the effect of varying inference batch sizes. This article significantly extends that idea, and proposes an interleaved execution approach for concurrent training and inference with low variability of inference latency; designs two novel optimization strategies, GMD and ALS, that meet inference latency and maximize training throughput while staying within a power budget; and reports comprehensive evaluation on 273,000+ configurations across 15 DNN workloads.

\vspace{-0.05in}
\section{Motivation}
\label{sec:motivation}

\begin{table*}[t]
\vspace{-0.1in}
\footnotesize
\caption{Practitioner's Matrix of Edge AI Scenarios and Solution Approaches.}
\label{tbl:PracMat}
\setlength{\tabcolsep}{1pt}

\begin{tabular}{L{1.5cm}|L{3.5cm}|L{1.8cm}|L{1.5cm}|L{1.2cm}|L{1.5cm}|p{1cm}}
\hline
\textbf{Workload} & \bf {Usecase} & \bf {Occurrence} & \bf {Runtime} & \bf {QoS}$^*$ & \bf {Approach} & \bf {Time to Sol.} \\
\hline\hline
\multirow{4}{1.3cm}{\bf Inference only}
& Continuous inferencing using single model, e.g., demand prediction from smart power meters & Frequent, predictable & Hrs to days & Latency & ALS & ~0.5-1.5hr \\
\cline{2-7}
& Various inference DNNs run on-demand, e.g., tracking a stolen vehicle using traffic cameras & Frequent, unpredictable & 10--60~mins & Latency & GMD & <10 min\\
\cline{2-7}
& Outlier inference tasks, e.g., voice assistant & Occasional, unpredictable & $<1$~min & Latency & MAXN mode & 0 \\
\hline
\bf Train only & Personalization/fine-tuning & One time, predictable & Few hrs& T'put & GMD & <10min \\
\hline
\multirow{2}{1.3cm}{\bf Inference + Train} & Continual learning & Frequent, predictable & <1 hr & Latency, T'put & ALS & <0.5-1.5hr \\
\cline{2-7}
& Inference+Federated learning background task & Unknown, unpredictable & 10--60~mins & Latency, T'put & GMD & <10 min\\
\hline
\multicolumn{7}{p{12cm}}{$^*$~\em QoS minimizes latency and/or maximizes throughput (t'put), with an added power budget as constraint.}
\end{tabular}
\end{table*}

\paragraph{Power Modes in Nvidia Jetsons}
\textit{Nvidia Jetsons} are the leading accelerated edge devices that are powerful enough to train DNN models using 1000+ CUDA cores and yet within 60W power load and a form factor no bigger than a novel~\cite{mlperf_inf}. 
They offer fine-grained \textit{power modes} that decide CPU core counts, and CPU, GPU and memory frequencies. The latest generation Jetson AGX Orin devkit, which we use in this article, has $12$ core counts, $29$ CPU frequencies, $13$ GPU frequencies and $4$ memory frequencies, resulting in $\approx 18,000$ power modes (Table~\ref{tbl:jetsonpowerspecs}). We can tune these for performance--power trade-offs~\cite{prashanthi2023sigmetrics}.

For instance, training \resnet on MAXN (highest performance) power mode takes $3.1$mins per epoch and $51.1$W of power while a lower power mode ($4$ cores, $422$MHz CPU, $115$MHz GPU, $665$MHz memory) increases the time to $25.6$mins per epoch but with a lower $14.7$W of power. Similarly, inference of \mobilenet with $bs=64$ on MAXN takes $102$ms per minibatch and $39.5$W of power; this grows to $1026$ms time but with a lower power of $14$W with the lower power mode. However, the non-linearity between performance and power~\cite{PowerTrain} means that selecting a power mode to meet the QoS goals is non-trivial. A wrong choice leads to high penalties.

\paragraph{Inferencing Minibatch Size}
The inferencing minibatch size ($bs$) also has a significant impact on power and performance.
E.g., \mobilenet inference on MAXN with $bs=1$ takes $18$ms per minibatch with a power of $20.9$W, while $bs=32$ takes $54$ms and $38.2$W of power -- a $3\times$ variation in time and $83\%$ in power. This also varies with DNNs. E.g., \bert inference on MAXN with $bs=1$ takes $66$ms per minibatch and $56$W of power while $bs=32$ takes $1.94$s and $61.8$W of power. Here, the time variation is much higher ($29\times$) but power varies less ($10.4\%$). This not only impacts inference latency and power budgets but also the training throughput in concurrent scenarios. Hence, tuning inference minibatch size is crucial.

\paragraph{Profiling Overheads and Predictions}
Profiling a power mode takes time, e.g., $2.4$--$102$s for a training workload and $120$ms--$6.4$s for inferencing on the Orin AGX, depending on the DNN. 
It is intractable to benchmark a large number of modes and $bs$ for every new DNN. E.g., it takes $>16$hrs to profile $\approx 4400$ (around $25\%$) power modes for ResNet training. 
Also, this will not scale for evolving workloads where models or arrival rates change, and for concurrent workloads where interference prevents extrapolation from standalone models.

The default power mode MAXN, while being the fastest, consumes maximum power and violates most power budgets, making it unsuitable for power-constrained deployments~\cite{PowerTrain}. In server GPUs, an explicit power cap can be set~\cite{zeus_nsdi23} and \textit{Dynamic Voltage and Frequency Scaling (DVFS)} kicks in to ensure that this is not violated. However, in Jetsons, there is no way to set a power cap, and therefore enabling DVFS does not help in meeting a power budget. For instance, \resnet inference with DVFS off and on consumed very similar powers of $22.15$W and $21.5$W on a given power mode ($8$ cores, $1651$MHz CPU, $522$MHz GPU, $3199$MHz memory) and violated the power budget of $15$W.

As we show later (\S~\ref{sec:results}), baselines to select a power mode, such as predicting the time and power for a power mode by training a Neural Network (\textit{NN250}) model~\cite{PowerTrain}, and selecting the best power modes from a pre-profiled set of modes (\textit{RND250}), are sub-optimal. 

\textit{Therefore, it is necessary to have a principled approach towards profiling as few power modes as possible to help identify the optimal configuration of power mode and minibatch size given a problem, even as the DNN workload changes.
}

\vspace{-0.05in}
\section{Review of Solution Approaches}
\label{sec:approach}

We consider $3$ types of workloads (Table~\ref{tbl:PracMat}) for which we design and evaluate possible solutions: \textit{Standalone training}, \textit{Standalone inference}, and \textit{Concurrent training and inference}.

\subsection{Approaches for Concurrent Training and Inference}
In the absence of hardware-assisted GPU sharing on the edge using MPS or MIG, there are three alternative approaches for concurrent training and inference: time sharing using native or managed interleaving and space sharing using GPU streams.

\begin{figure}[t]
\vspace{-0.1in}
\centering
\includegraphics[width=0.50\columnwidth]{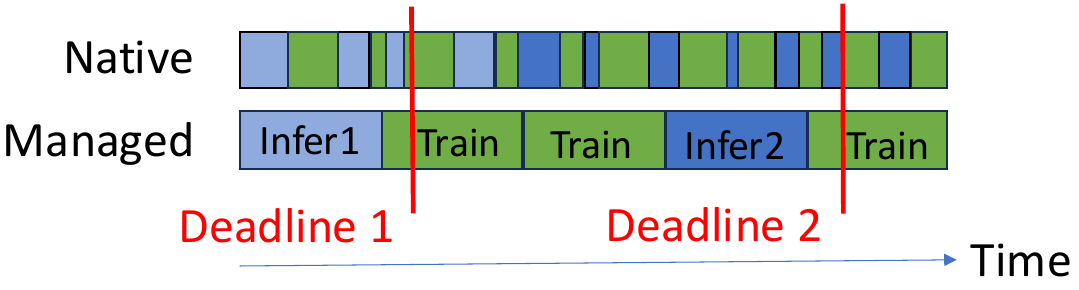}
\caption{Native and Managed Interleaving. \textit{Infer1} for Native misses \textit{Deadline1}}
\label{fig:intnatman}
\end{figure}

\textit{Native interleaving} is the \textit{de facto} scheduling mechanism for concurrent workloads.
The training and inference processes run simultaneously and their \textit{time-sharing} of the GPU is decided by the GPU scheduler~\cite{edge_gpustreams, multiinf_taas23}. Interleaved execution happens at a fine \textit{kernel-level granularity}, oblivious to any user QoS goals (Figure~\ref{fig:intnatman}, top). 

Another approach is to use \textit{CUDA streams}~\cite{streams} to run training and inference on separate streams using multi-threading. The GPU scheduler runs them simultaneously in a \textit{space-shared} manner on the SMs when there are adequate GPU resources. 
We can also set two priorities (high and low) for each stream. These are considered whenever a block finishes execution and the next block needs to be scheduled~\cite{tx2scheduling_rtss}.

Lastly, we propose \textit{managed interleaving}, discussed more in \S~\ref{sec:strategies}. Our application-level scheduler (\papertitle) explicitly executes only one DNN at a time, at the \textit{granularity of minibatches} (Figure~\ref{fig:intnatman}, bottom).
We intelligently modulate the number of training minibatches to execute, followed by an inference minibatch with a tuned minibatch size, and again run training minibatch(es) and inferencing, and so on. 
This explicit scheduling is based on the inference request arrival rate, the inference latency deadline, and the execution duration, which depends on selected the power mode. For a steady inference arrival rate and a static pair of training/inferencing models, we can find a static solution; but if either model changes the solution needs to retune the knobs.
Managed interleaving is also \textit{time-sharing}, but at a minibatch granularity and controlled at the application level by \papertitle.

\subsection{Performance Comparison Summary}
We summarize the performance of these three approaches to contrast their behavior, and justify our design choice. We simultaneously execute inference and training of \mobilenet using these three approaches for $10$ diverse problem configurations (\textit{Cfg}), defined by the inference arrival rate ($40$--$120$RPS), the inference latency budget ($600$--$1200$ms) and power budget ($22$--$40$W).
We decide the execution settings such as power mode and inference minibatch size using our GMD strategy, discussed in \S~\ref{sec:strategies}. For streams, the inference workload is sent to a high-priority stream, while training is sent to a low-priority stream.
Each configuration is run for $200$ training minibatches, and runs for $1$--$3$mins.

\begin{figure}[t]
\vspace{-0.1in}
\centering
\includegraphics[width=0.95\columnwidth]{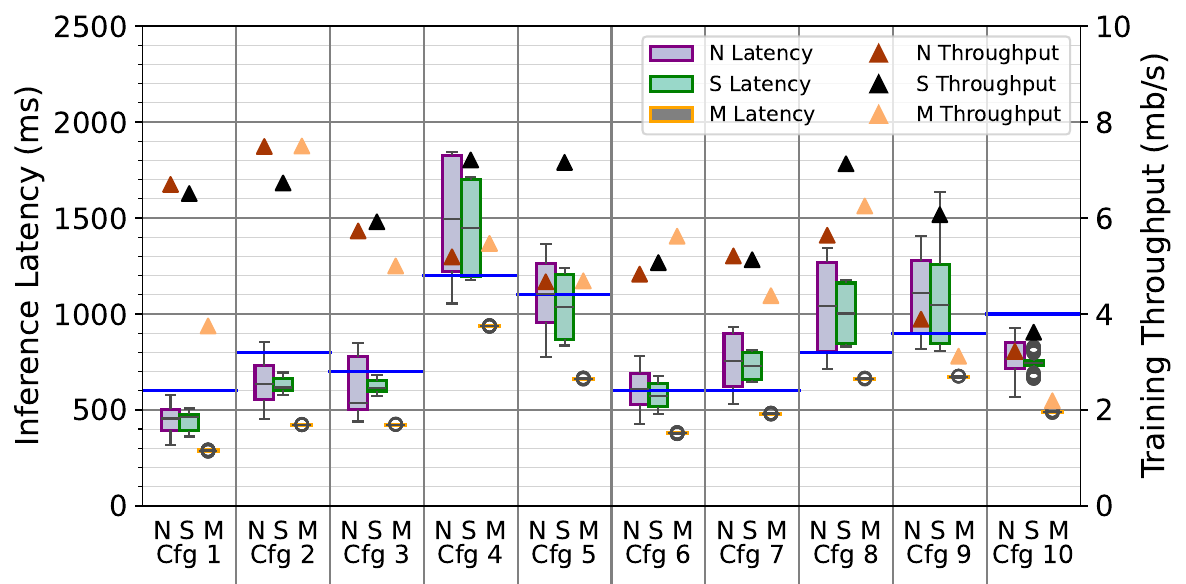}
\caption{Inference latency \textit{(left Y axis, box plots)} and training throughput \textit{(right Y axis, markers)} for concurrent workloads executed using \textit{Native Interleaving (\textbf{N})}, \textit{Streams (\textbf{S})} and \textit{Managed Interleaving (\textbf{M})} for $10$ problem configs.}
\label{fig:intconc}
\end{figure}

Figure~\ref{fig:intconc} shows a box plot of the inference latency per minibatch (queuing time + execution time) on the left Y axis, and training throughput as markers on the right Y axis for the different workload configurations on X axis.
The blue horizontal line indicates the inference latency budget.

We see that the inference time for native interleaving (N) is highly variable, with the third quartile (Q3) often violating the latency budget, and occasionally even the median (Q2) exceeding it. This is due to fine-grained switching of executions between the training and inferencing models. For streams (S), although the median latency is slightly lower than native, we still see wide variability leading to latency violations. Previous work on concurrent inferencing using streams~\cite{multiinf_taas23, tx2scheduling_rtss} report similar behavior due to its non-deterministic resource blocking.
In contrast, our managed interleaving (M) has a tight distribution and always stays within the inference latency deadline due to careful switching between the model executions at minibatch boundaries.
This leads to predictable completion times.

But the training throughput may suffer with managed interleaving (e.g., Cfg~1) compared to native due to executing an \textit{integral} number of training minibatches between inferences. This causes some idle time within an optimization window. Similarly, streams occasionally achieve a higher training throughput due to resource sharing. Streams are well suited for configurations where training and inferencing are both throughput-sensitive and latency variability is not a concern. However, our focus in this article is to achieve a \textit{stable inference latency} within the given latency budget while \textit{maximizing training throughput}. Hence, we use managed interleaving in the rest of the article for concurrent workloads. 
Since it operates at a coarser granularity, understanding the power and performance of training and inferencing models separately can help solve concurrent training and inference. This reduces the complexity of the problem from $m \times n$ to $m+n$ model combinations.

\vspace{-0.05in}
\section{Problem Formulation}
\label{sec:problem}

Here, we formally define the scheduling problem as an optimization problem. Based on the workload, there are 3 variants: \textit{standalone training}, \textit{standalone inference} and \textit{concurrent training and inference}. 

A summary of notations is in Table~\ref{tbl:notation}. 

\begin{table}[t]
\vspace{-0.1in}
\centering
\setlength{\tabcolsep}{1.5pt}
\renewcommand{\arraystretch}{0.9}
\def\thickhline{\noalign{\hrule height1pt}}
\small
\caption{Notations Used}
\label{tbl:notation}
\begin{tabular}{C{1cm}|C{5cm}|C{3.5cm}}
\hline
    \textbf{Symbol} & \textbf{Description} & \textbf{Formula} \\
  \thickhline
 $\mathbb{M }$ & Set of device power modes & \\
    \hline
    $pm$ & power mode & \\
    \hline 
    $\widehat{p}$ & Power budget (W)& \\
  \thickhline
    $\mu_{tr}$ & Training model & \\ 
    \hline
    $t_{tr}$ & Train minibatch time (s) & $t_{tr}=\mathcal{F} (\mu_{tr}, pm)$ \\ 
    \hline
    $\theta_{tr}$ & Training throughput (minibatches/s) & $\theta_{tr}=1/t_{tr}$ \\
    \hline
    $p_{tr}$ & Training power load & \\
     \hline
    $\tau_{tr}$ & No. of training minibatches & \\
     \thickhline
    $\mu_{in}$ & Inf model & \\
    \hline
    $\beta_{in}$ & Inf minibatch size & \\
    \hline
    $\alpha_{in}$ & Inf arrival rate (RPS) & \\
    \hline
    $p_{in}$& Inference power load (W) & \\
    \hline
    $qt_{in}$ & Peak queueing time (s) & $qt_{in} = (\beta_{in}-1)/\alpha_{in}$ \\
    \hline
    $t_{in}$ & Inf minibatch time (s) & $t_{in}=\mathcal{G} (\mu_{in}, \beta_{in}, pm)$\\
    \hline
     $\lambda_{in}$ & Peak minibatch infer. latency (s/req) & $\lambda_{in}=qt_{in} + t_{in}= (\beta_{in}-1)/\alpha_{in}+t_{in}$\\
    \hline
    $\widehat{\lambda}$ &Inf latency budget (s/req) & \\
    \hline
\end{tabular}
\end{table}

\paragraph{Standalone Training} 
Here, only a single training workload $\mu_{tr}$ is running, and the user provides a power budget $\widehat{p}$. 
The goal is to select a power mode $pm \in \mathbb{M}$ that maximizes the training throughput $\theta_{tr}$ while the power load during training $p_{tr}$ stays within the given power budget.
\setlength{\jot}{0pt}
\begin{gather*}
\text{Given} \ \mu_{tr},\ \sel \ pm \in \mathbb{M} \\
\arg\max \ \theta_{tr} \  \subto \ p_{tr} \leq \widehat{p}
\end{gather*}

\paragraph{Standalone Inference} 
Here, the user specifies a power budget $\widehat{p}$ and a latency budget $\widehat{\lambda}$. The input arrival rate $\alpha_{in}$ of requests for inferencing is assumed to be constant. Since inference is the only workload,  we need to select a power mode $pm$ and inference minibatch size $\beta_{in}$ such that we minimize the inference latency $\lambda_{in}$, stay within the latency budget, and ensure that the power load $p_{in}$ during inferencing remains within the power budget. 

\setlength{\jot}{0pt}
\begin{gather*}
\text{Given} \ \mu_{in} \ \text{and} \ \alpha_{in}, \ \sel \ pm \in \mathbb{M} \ \text{and} \ \beta_{in}\\
\arg\min \ \lambda_{in} \ \subto \ \lambda_{in} \leq \widehat{\lambda} \ \text{and} \ {p_{in}} \leq \widehat{p}
\end{gather*}

\begin{figure}[t]
\vspace{-0.1in}
\centering
\subfloat[Large minibatch size]{
    \includegraphics[width=0.45\columnwidth]{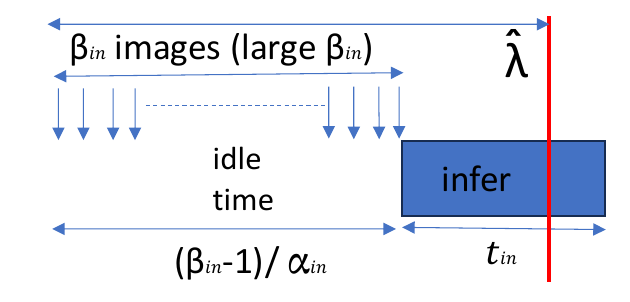}
    \label{fig:infex2}
  }~~
\subfloat[Small minibatch size]{
    \quad \includegraphics[width=0.35\columnwidth]{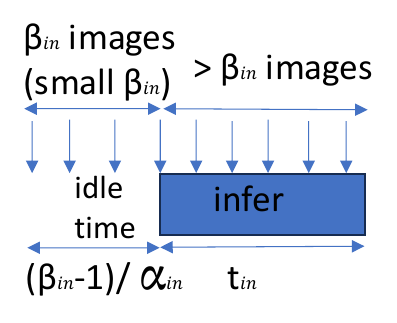} \quad 
    \label{fig:infex3}
  }
  
\caption{Examples of inference latency violation }
\label{fig:infexs}
\end{figure}

In Figure~\ref{fig:infexs}, we illustrate an inference workload and the conditions that need to be met. Initially, inference is idle, waiting for a minibatch size of $\beta_{in}$ to accumulate. The images arrive at a rate of $\alpha_{in}$, leading to a queuing time of $\frac{(\beta_{in}-1)}{\alpha_{in}}$. This minibatch takes time $t_{in}$ to execute, during which images for the next minibatch are arriving.

There are two counterpressures at play here. In Figure~\ref{fig:infex2}, if the minibatch size is increased beyond a limit, the queuing and execution times increase, leading to a latency violation as shown. On the other hand, if the minibatch size is too small, as in Figure~\ref{fig:infex3}, the processing time can become too high, resulting in more requests queuing up than can be processed, i.e., the inference rate cannot keep up with the arrival rate, leading to unbounded queuing and latency violations. \textit{The goal, therefore, is to find the smallest inference minibatch size that satisfies the latency and power constraints.}

\paragraph{Concurrent Training and Inference}
\label{sec:interleaved}

\begin{figure}[t]
\vspace{-0.1in}
\centering
\subfloat[Minibatch size that only maximizes training minibatches]{
    \quad \includegraphics[width=0.45\columnwidth]{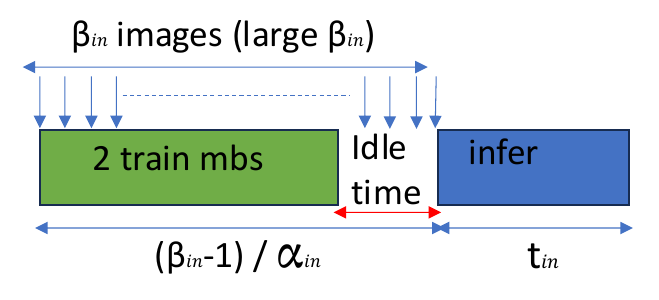}\quad 
    \label{fig:infex4}
  }\qquad 
\subfloat[Minibatch size that also minimizes inference latency]{
    \quad \includegraphics[width=0.35\columnwidth]{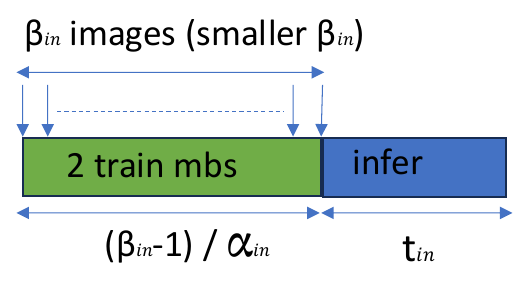}\quad 
    \label{fig:infex5}
  }
\caption{Examples of Managed Interleaving}
\label{fig:intexs}
\end{figure}

This workload consists of both training and inference models, $\mu_{tr}$ and $\mu_{in}$. The user specifies the peak power load $\widehat{p}$ and inferencing latency $\widehat{\lambda}$. We need to select a power mode and inference minibatch size to maximize training throughput $\theta_{tr}$ while staying within the latency and power budgets. A secondary goal is to minimize the inference latency $\lambda_{in}$.
\setlength{\jot}{0pt}
\begin{gather*}
\text{Given} \ \mu_{tr}, \ \mu_{in} \ \text{and} \ \alpha_{in},\  \sel \ pm \in \mathbb{M}, \ \beta_{in} \ \text{and} \ \tau_{tr}\\
Primary: \arg\max \ \theta_{tr} \ \subto \ \lambda_{in} \leq \widehat{\lambda} \ \text{and} \ {p} \leq \widehat{p}\\
Secondary: \arg\min \ \lambda_{in}
\end{gather*}

We use managed interleaving of training and inference, with $\tau_{tr}$
training minibatches executed for every inference minibatch (Figure~\ref{fig:intexs}).
Increasing the minibatch size $\beta_{in}$ increases the inference latency but also allows a larger number of training minibatches to be executed since it idles longer for requests to queue up. Since the inference time increases sub-linearly with the minibatch size, the growth in inferencing latency is mitigated.
But the inference latency budget should not be violated. So, we pick the largest $\beta_{in}$ that meets the latency. As a secondary goal, we aim to minimize the inference latency; so, if two $\beta_{in}$ sizes result in the same number of training minibatches, we pick the smaller one (Figure~\ref{fig:intexs}).

\paragraph{Discussion}
In this article, we initially focus on running one training and one inference model, concurrently, at a static inference input rate. We then extend our study to dynamic inference arrival rates and two concurrent inferences. Our methods can be extended to 3 or more models easily, but we limit our study to two models at a time in this article, and leave a higher level of concurrency to future work. We do not tune any DNN training hyper-parameters such as minibatch size since, unlike inference, these can affect the accuracy achieved. So, the term minibatch size in this article always refers to inference minibatch size.

\begin{figure}[t]
\vspace{-0.1in}
\centering
\includegraphics[width=0.9\columnwidth]{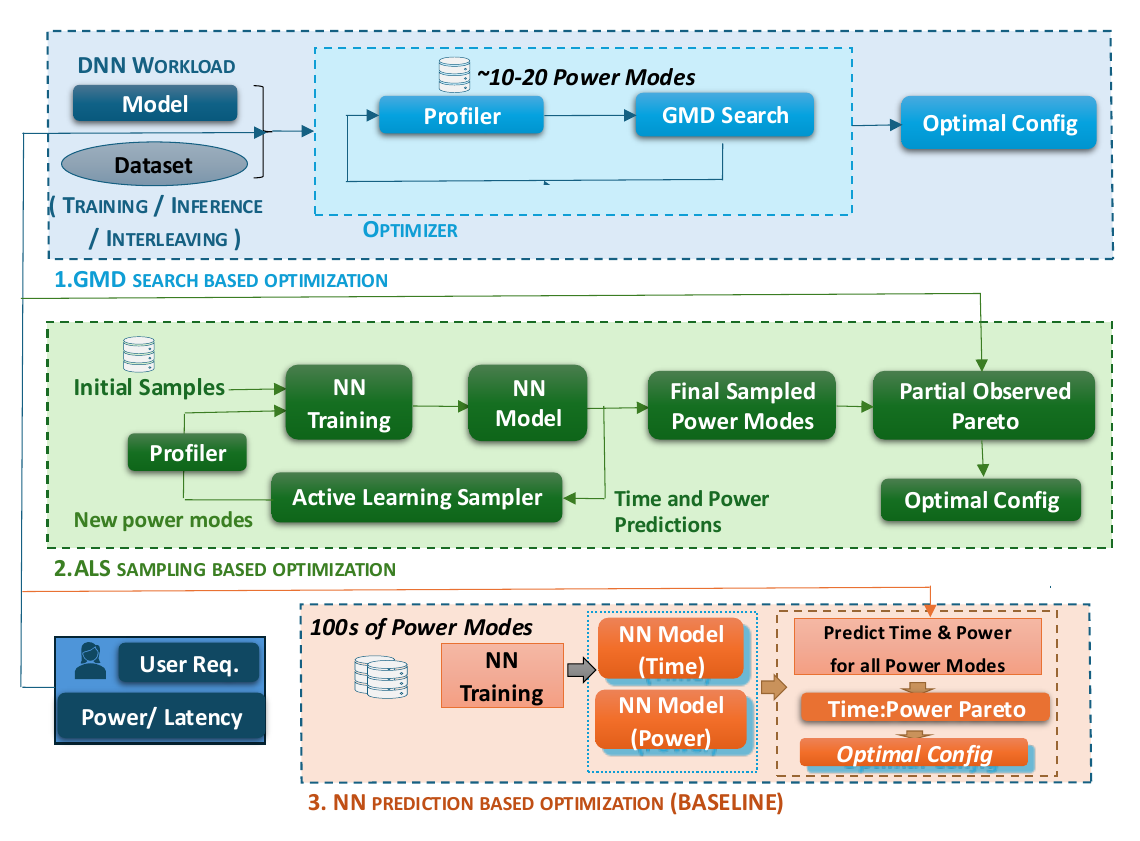}
\caption{Workflow to solve scheduling problem using \textit{\papertitle} (GMD and ALS, top) and \textit{NN baseline} (bottom).}
\label{fig:workflow}
\end{figure}

\section{Proposed Strategies}
\label{sec:strategies}
Throughout this article, we use the term \textit{workload} to refer to DNN model(s) (training or inference or a concurrent pair), and \textit{problem configuration} to refer to user-specified requirements such as inference arrival rate, latency budget and power budget for the workload.

We propose two strategies to solve the optimization problem, ALS and GMD, as described below. 
Their merits for addressing the different workload scenarios are summarized in Table~\ref{tbl:PracMat}.
GMD profiles $10$--$15$ power modes to quickly arrive at a solution within $5$--$10$mins for each problem configuration. In contrast, ALS profiles $50$--$150$ modes, which takes $\approx1.5$hrs, but can generalize to other problem configurations of the workload, e.g., with varying power and latency budgets, and arrival rates.

Also, if the workload or the edge devices changes, then again both ALS and GMD require fresh profiling. They are used within \papertitle to execute a given workload, as illustrated in Figure~\ref{fig:workflow}. 

\subsection{Gradient-descent based Multi-Dimensional Search (GMD)}
We have a 4D solution space of CPU/GPU/memory frequencies and CPU cores, which together decide the power mode.
Our broad solution approach in GMD is to profile the workload for an initial power mode, and use that knowledge to prune the search space and decide the next power mode to select and profile.  We repeat this until we find a power mode that meets the problem constraints and maximizes/minimizes the optimization goal within the space we searched in. The intent is to perform as few profiling runs as possible to reduce the time to solution. That said, these runs are also valid workload executions and their outputs can be used by the application, though the time and power constraints may not be met.
Once a power mode has been profiled for a DNN, it can be reused in future problem configurations too.

\begin{figure}[t]
\vspace{-0.1in}
\centering
\subfloat[Simple Binary Search. X axis has GPU freq. (outer) and CPU cores (inner). Y axis has CPU freq. (outer) and mem. freq. (inner).]{%
    \qquad \qquad \includegraphics[width=0.4\columnwidth]{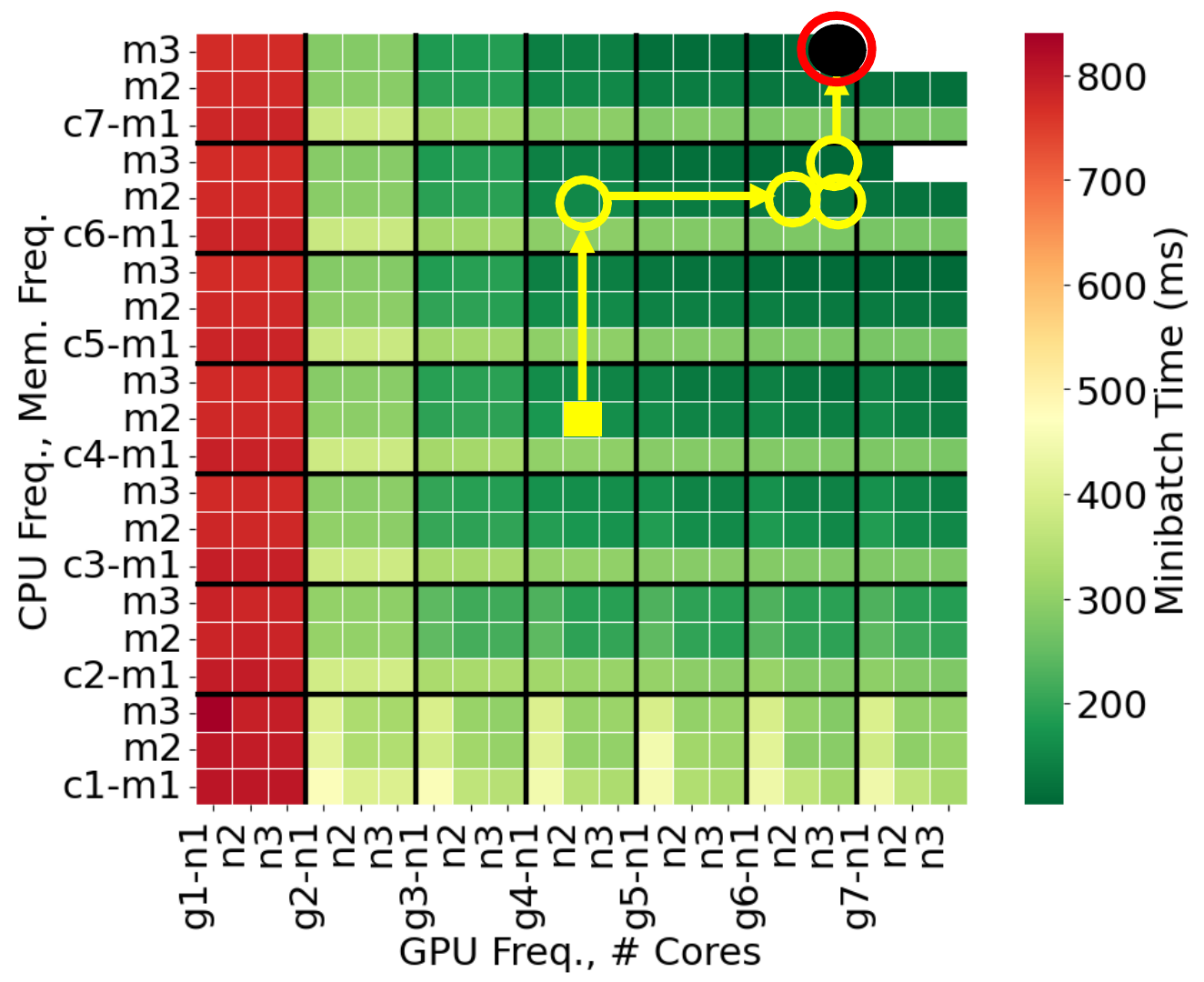}\qquad \qquad 
    \label{fig:Binary_Search_Example}%
}\qquad 
\subfloat[GMD]{%
    \raisebox{0.3in}{\includegraphics[width=0.3\columnwidth]{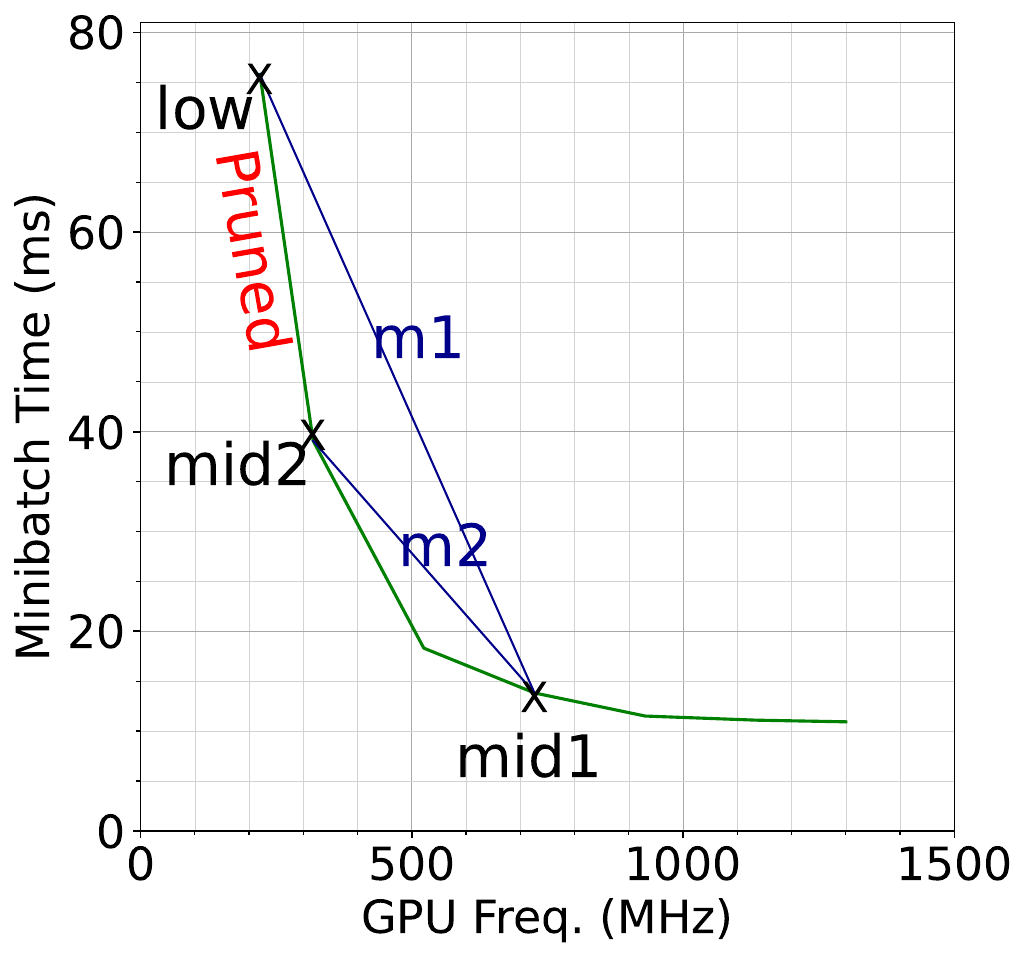}}%
    \label{fig:Binary_Search_Curve}%
  }%
\caption{Comparing the search behavior of Binary Search and GMD}%
\end{figure}

A key challenge is to decide which power mode to start at and in which direction to search in this solution space.
A \textit{simple binary search} starts at the center of all dimensions and visits each dimension in a round-robin fashion. In each iteration, it prunes half of the remaining space to be visited along a dimension based on whether the profiled power is under or over the power budget.
This returns a solution in $\log{n}$ profiling trials, where $n = cores \times cpuf \times gpuf \times memf$, shown  across 4 dimensions (4D) of the power modes in Figure~\ref{fig:Binary_Search_Example} (X and Y axis are nested resource dimensions). The heatmap colors indicate minibatch latency. The yellow square is the starting point, yellow lines and circles give the search trajectory, the red circle is the returned solution, and the black disc is the optimal solution.
But, by visiting the dimensions in a fixed round-robin order, this strategy incorrectly prunes out viable candidates. \textit{We improve upon this in GMD by using domain knowledge to decide the next dimension to visit.}

\begin{figure}[t]
\centering
\vspace{-0.1in}
\subfloat[Train Time vs. GPU Freq.]{
    \includegraphics[width=0.40\columnwidth]{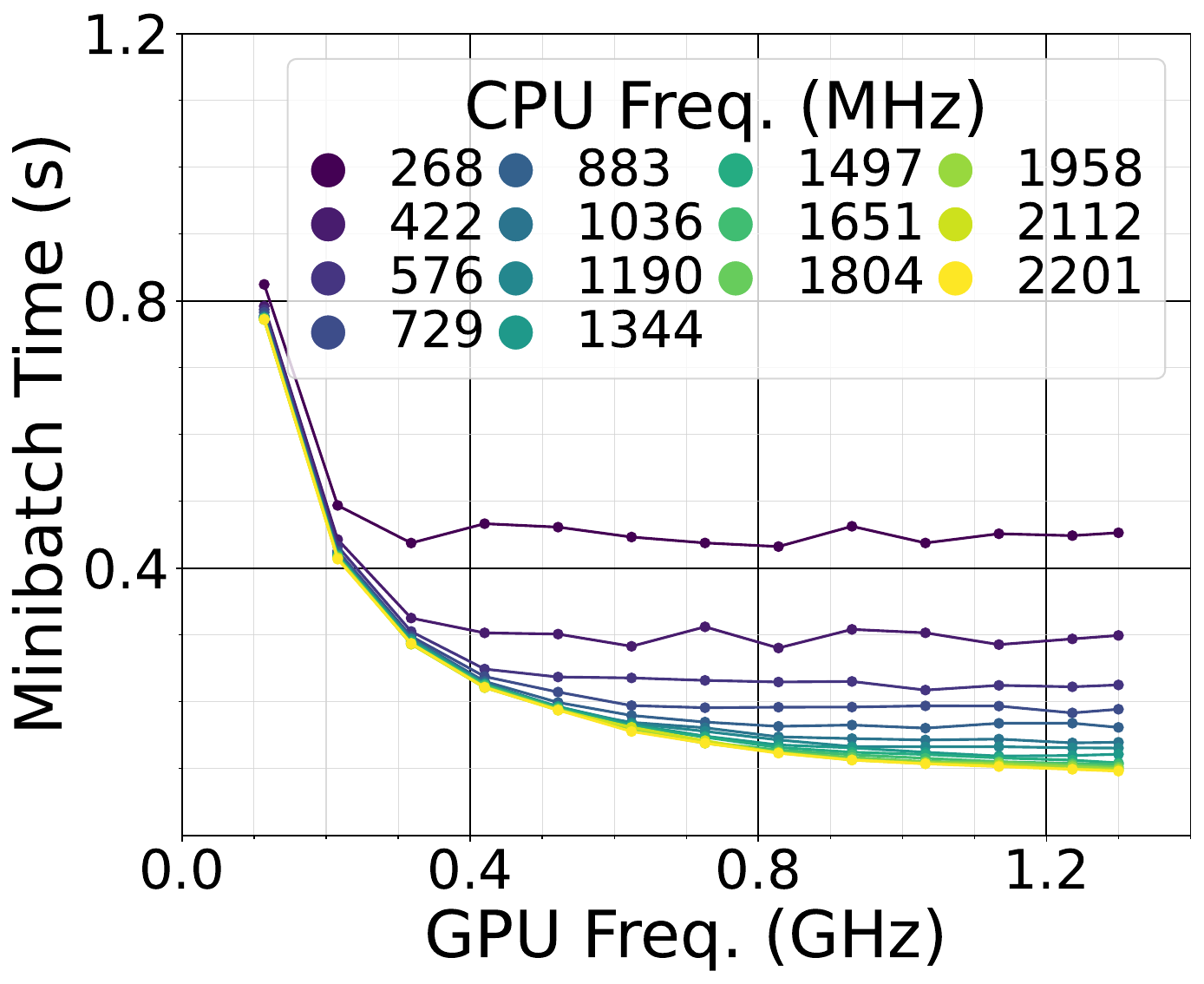}
    \label{fig:domain_aware_time}
  }\quad 
\subfloat[Power vs. GPU Freq.]{
    \includegraphics[width=0.40\columnwidth]{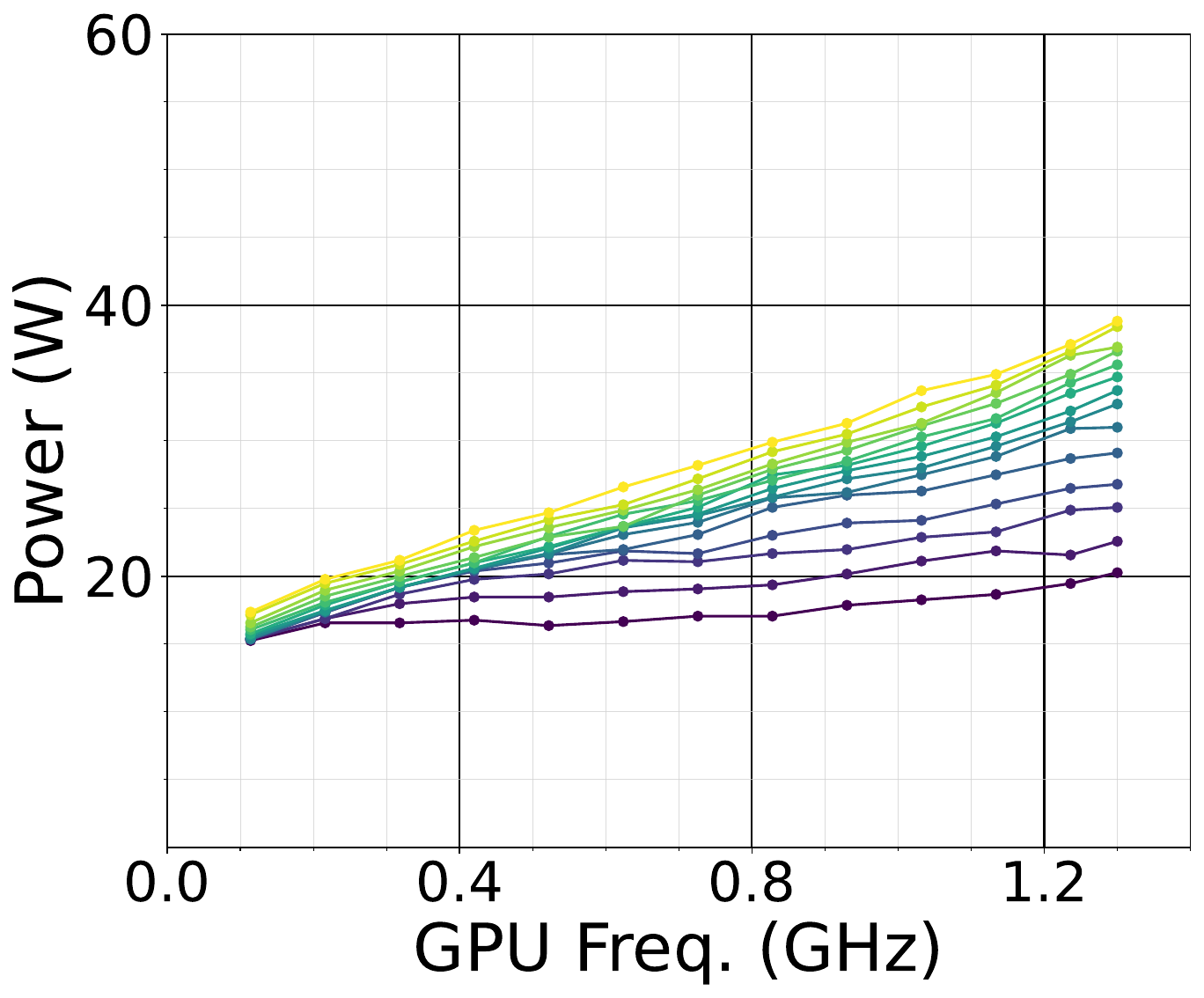}
    \label{fig:domain_aware_power}
  }

\caption{MobileNet training time and power vs. GPU and CPU frequencies \textit{(Cores=12, Mem Freq=2133)}.}
\label{fig:domain_aware}
\end{figure}

\begin{figure*}[t]
\vspace{-0.1in}
\centering
\subfloat[GMD]{%
    \includegraphics[width=0.75\textwidth]{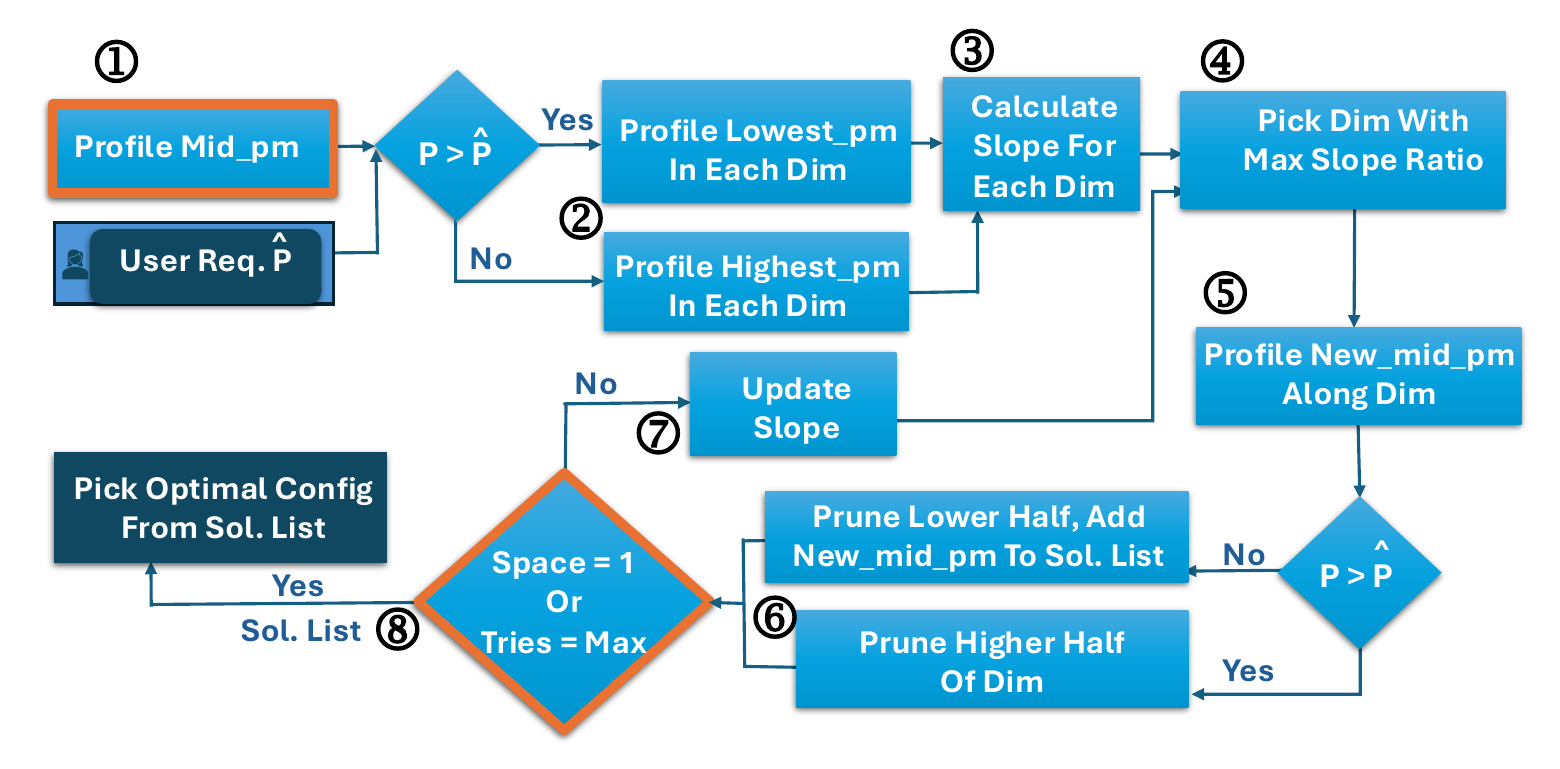}%
    \label{fig:GMDtrain}%
}\\
\subfloat[ALS]{%
    {\includegraphics[width=0.75\textwidth]{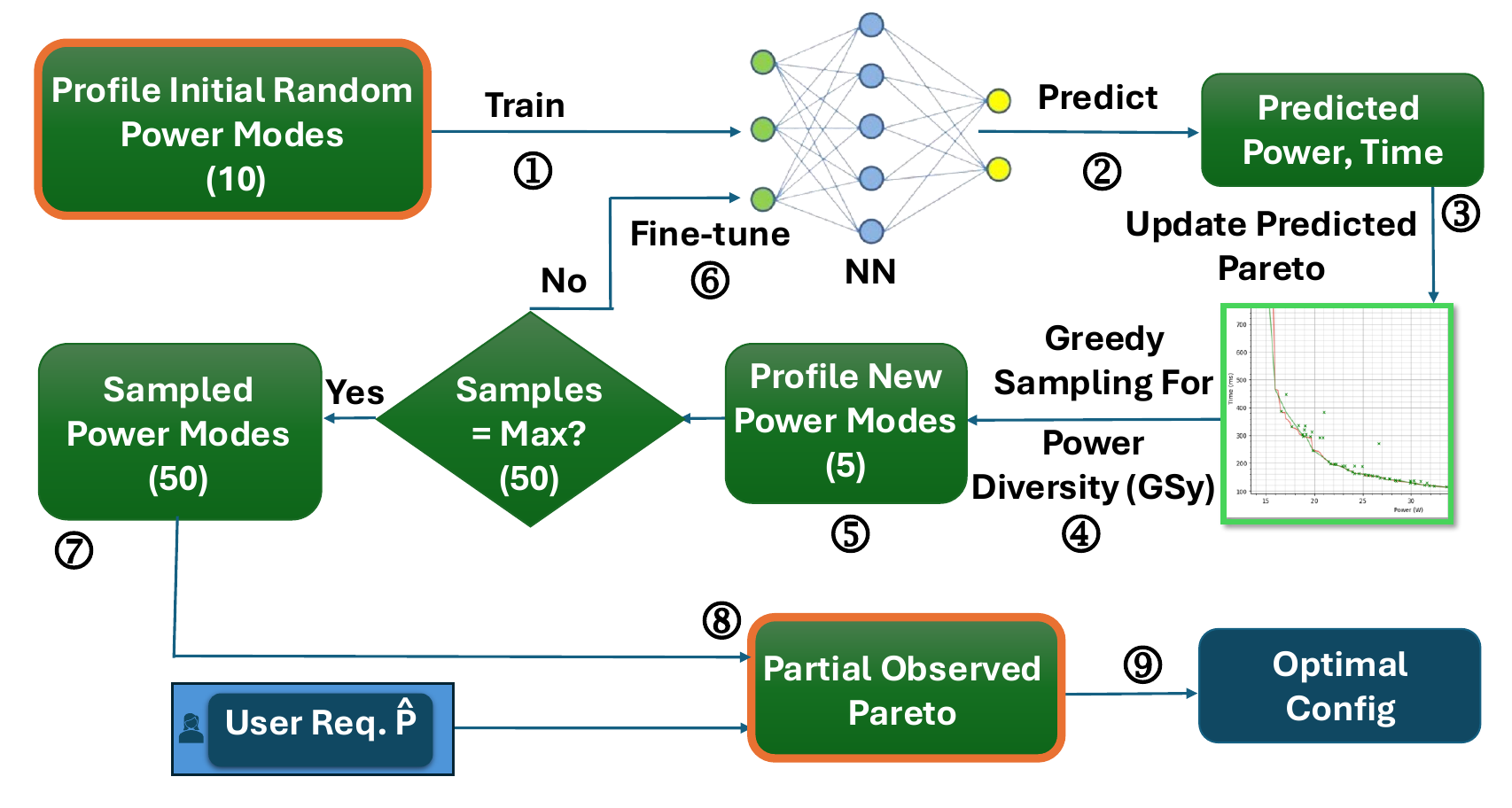}}%
    \label{fig:ALStrain}%
  }%
\vspace{-0.15in}
\caption{GMD and ALS Strategies for Standalone Training} 
\label{fig:GMDALSworkflow}

\vspace{-0.15in}
\end{figure*}

\subsubsection{GMD Intuition} We plot the variation in minibatch training time and power load during training against GPU frequency, for various CPU frequencies and a static core count ($12$) and memory frequency ($2133$MHz). 
The training time is non-linear with GPU frequency (Figure~\ref{fig:domain_aware_time}). It initially drops sharply as the GPU frequency rises and then saturates. But power steadily increases with GPU frequency (Figure~\ref{fig:domain_aware_power}). We confirm that this trend holds for diverse DNN models, and only the slope of the time and power varies across dimensions and workloads. We leverage this knowledge in GMD. 

Specifically, we use the insight that not all resource dimensions contribute equally to power and performance, and giving them equal chance to influence the power mode (like binary search above) is not efficient for profiling. Instead, we find the \textit{rate of change} in time and power when we change a dimension's value (keeping other dimensions the same). The \textit{time slope} and \textit{power slope} give us this, i.e., for a unit change in the dimension's value, what is the $m^{time}$ change in time and $m^{pow}$ change in power. 

We then take the ratio between these two slopes, $\rho=\frac{m^{time}}{m^{pow}}$, to understand the relative influence of each dimension on the time vs. power. This helps us decide at each step of the search, which dimension of the power mode should be changed to give us the steepest drop in time or power, and most quickly get us close to the optimal solution. 

This prioritized search should help converge to a suitable solution with fewer profiling runs than binary search. Since GMD uses the profiled slope for each workload, it can adapt based on which dimension that workload is most sensitive to, unlike static sampling baselines where there is no workload-specific change. We also tune batch size for inference using GMD, as yet another dimension. GMD needs to be rerun for each problem configuration of training/inference/concurrent DNN workloads. 

\subsubsection{GMD for Standalone Training} 
We illustrate the GMD algorithm in Fig.~\ref{fig:GMDtrain}, provide its pseudo-code in the Appendix, Algorithm~\ref{algo:gmd}, and describe them here.
We first profile the DNN training model on the edge device using the \textit{midpoint power mode} $(mid1)$ that has all \textit{resource dimensions} set to their midpoint values (Fig.~\ref{fig:GMDtrain}~\circled{1}, Algorithm~\ref{algo:gmd} L~\ref{algo_gmd:init:start}--\ref{algo_gmd:init:end}), e.g., $mid1$=\texttt{8c/\-1344MHz/\-727MHz\-/\-2133\-MHz} as CPU core count/CPU freq/GPU freq/Mem freq for Orin AGX (Table~\ref{tbl:powermodes_exp}, Figure~\ref{fig:Binary_Search_Curve}). 
This profiling returns the training time per minibatch and the power load for this mode.

We then select $4$ more power modes, one per dimension, in a direction guided by the power consumed by the midpoint mode -- if the power ($p_{mid1}$) is \textit{over budget ($\widehat{p}$)}, the $4$ modes picked will each set one dimension to its \textit{lowest value} and retain values for the other three; and if \textit{below budget}, use the \textit{highest value} (\circled{2}, L~\ref{algo_gmd:mid:start}--\ref{algo_gmd:mid:end}). E.g., if $p_{mid1} >\widehat{p}$ for Orix AGX, we select power modes:
\begin{itemize}[noitemsep,topsep=0pt]
    \item low1$_{core}$ \,\,= {\tt \textbf{4c}/1344MHz/727MHz/2133MHz}
    \item low1$_{cpuf}$ \,= {\tt 8c/ \textbf{422MHz}/727MHz/2133MHz}
    \item low1$_{gpuf}$ \,= {\tt 8c/1344MHz/\textbf{115MHz}/2133MHz}
    \item low1$_{memf}$ = {\tt 8c/1344MHz/727MHz/ \textbf{665MHz}}
\end{itemize}
Else, if $p_{mid1} \leq \widehat{p}$, we select power modes high1$_{core}$=\texttt{\textbf{12c}/\-1344MHz/\-727MHz/\-2133MHz}, and similarly for high1$_{cpuf}$, high1$_{gpuf}$ and high1$_{memf}$.

Using the midpoint combined with each of the $4$ additional (highest/lowest) modes, we fit $4$ sets of time and power slopes (\circled{3}).
E.g., in Figure~\ref{fig:Binary_Search_Curve}, for the GPU frequency dimension, we start with points $mid1$ (727MHz) and low1$_{gpu}$ (115MHz), and calculate their slope $m1^{time}_{gpuf}$ for the minibatch training time; and similarly, the slope $m1^{pow}_{gpuf}$ for the power load profile.

For each dimension, we get the ratios of the time slope to the power slope, e.g., $\rho_{gpuf}=\frac{m^{time}_{gpuf}}{m^{pow}_{gpuf}}$. 
We pick the dimension with the highest slope ratio ($\rho_{max}$) (\circled{4})  for the next \textit{search and prune} step. This uses the insight that some dimensions contribute more to the performance--power tradeoff for each workload. The dimension with the highest $\rho$ slope ratio gives the steepest drop in time for the least increase in power. So, exploring this dimension first can lead to a solution faster. For instance, memory frequency is the dimension with the highest slope ratio for \resnet training at $\rho_{memf}=8.93$, compared to $\rho_{core}=0$, $\rho_{cpuf}=1.59$ and $\rho_{gpuf}=1.8$. So changing the memory frequency gets us the most change in training time for a unit change in power.
We also observe that the slopes and their ratio can vary as GMD progresses and the dimension values for the power modes change, as well as across workloads.

We pick the new middle value ($mid2$) along this dimension to profile next, either from the lower or the higher half, i.e., whether the previously chosen $4$ modes were in the lower or higher half; the other dimension values stay the same (\circled{5}). If the profiled power mode $p_{mid2}$ satisfies the power budget, we add it to the list of candidate solutions. We prune the space along this dimension based on the power (\circled{6}), and then update the slope for this dimension (\circled{7}). 
E.g., in Figure~\ref{fig:Binary_Search_Curve}, we profile $mid2$ and eliminate the region $low1$ to $mid2$ (shown in red text), based on the fact that its power $p_{mid2} <\widehat{p}$ and the optimal solution is unlikely to lie here. This is because power increases monotonically along all dimensions. Therefore, if $p_{mid2}<\widehat{p}$, and we have 
$p_{low1}<p_{mid2} \forall low1<mid2$; hence $p_{low1}<\widehat{p}$.

We update the slope to calculate $m2$ between $mid1$ and $mid2$, as this is a closer approximation of the impact of GPU frequency on time and power in our new region of interest (L~\ref{algo_gmd:profile:start}--\ref{algo_gmd:profile:end}). Also, we use a thresholding logic to detect and omit cases where a negligible decrease in power artificially inflates the slope. 

Once all values in this dimension with the best slope have been exhausted, we move to the dimension with the next highest slope and repeat the process, till we exhaust the search space or run out of the profiling budget (\circled{8}, L~\ref{algo_gmd:prune:start}--\ref{algo_gmd:prune:end}).
The \textit{profiling budget} is a hyper-parameter that is set to $10$ tries since we observe that in most cases, the search space is pruned within 8 tries. 
Finally, we return the best among the candidate solutions. 

\subsubsection{Changes for Standalone Inference} 

In the case of standalone inference, we also need to tune the \textit{inference minibatch size}. One possibility is to treat the minibatch size like the other resource dimensions to explore its values. But we observe that it routinely has a high slope ratio, resulting in smaller minibatch sizes getting incorrectly pruned early on and resulting in a suboptimal solution with high latency. Therefore, we treat it as a special dimension that is visited differently from the others. 

We run the first iteration of GMD with $bs =  1$ to start with the minimal inference latency (\circled{A} in Fig.~\ref{fig:GMDinfer} in Appendix). 

\textit{Backtracking:} During training, GMD always finds a solution because power is the only constraint. However, inference has additional constraints on the latency budget at a given arrival rate. So, GMD may not find a solution with $bs =  1$ due to latency violations. If no solution is found for $bs =  1$, then we \textit{backtrack} as follows.

Among the power modes visited in the first iteration, we identify those that satisfy the power budget, but violate the latency because the inference rate cannot keep up with the arrival rate. Our experiments show that smaller $bs$ have a sublinear growth in time; so increasing $bs$ helps meet the arrival rate (\circled{B}). We try these power modes on the larger $bs$ values (\circled{C}), sorted by increasing latency, till we find a solution. The profiling budget is set to \textit{11 power modes} in this case to allow $1$ extra power budget for backtracking.

\subsubsection{Changes for Concurrent Workload}  
We can maximize the training throughput goal for concurrent workloads by maximizing the inference $bs$. As described in \S\ref{sec:problem}, this is due to the sublinear growth in inference latency which gives more time for training without violating the inference latency budget.

So, we change GMD to set the inference minibatch size to the largest candidate size we use, $bs =  64$ (\circled{D}, Figure~\ref{fig:GMDconcur} in Appendix). Also, the profiling budget is set to \textit{15 modes} here. to allow $5$ extra power modes for \textit{backtracking} and for the \textit{initial branch and bound}, described below. We allocate a larger budget because choosing the wrong batch size in the branch and bound step can lead to severely suboptimal solutions.

\paragraph{Branch and Bound}
Before we start the GMD search, we check if the fastest power mode, MAXN, can meet the inference latency budget for this $bs$. If not, we eliminate this $bs$, since every other slower power mode will have a higher execution time than MAXN and hence violate latency. We move to the next lower $bs$ and try this until we reach a $bs$ that meets the constraint at MAXN (\circled{E}). We limit our multi-dimensional search to this $bs$.

A key difference during this GMD search is that we consider the slope ratios of the \textit{dominant} workload among inferencing and training. At each step, by profiling for one power mode, we determine if inferencing or training uses more power, and this is the dominant workload for that step. This is because power is a system-wide constraint which is determined by the higher of the two powers. So we use the slope ratios of the dominant workload to decide which dimension to visit and prune. 

\paragraph{Backtracking}
If no solution is found in the first retained $bs$, we try to identify candidate solutions with smaller $bs$ (\circled{F}). Unlike for standalone inference, we eliminate those that cannot keep up with the arrival rate -- moving to a smaller $bs$ will not find us a solution because inference rate further decreases and will not satisfy the arrival rate.

Among those left, we move to the next lower $bs$ and try candidates sorted by increasing latency.

\subsection{Neural Network Baseline (NN)}
\label{sec:nn}
We first describe a NN baseline, proposed earlier~\cite{PowerTrain}, followed by our ALS strategy that improves upon this.
Neural Networks~\cite{mcculloch1943logical} can model complex non-linear relationships. We adapt our prior NN approach in PowerTrain~\cite{PowerTrain} for the current problem. We train a NN model to predict the minibatch training (or inferencing) time, and another to predict the power load. These are trained over profiling results from various power modes for a DNN (and inference minibatch size).

The NN has $4$ dense layers with $256, 128, 64,$ and $1$ neurons, designed based on a hyper-parameter search using \textit{RandomSearchCV}. ReLU is the activation function for the first $3$ layers and linear for the last. Adam optimizer is used with a learning rate of $0.001$. We define a custom Mean Average Percentage Error (MAPE)-based loss function that penalizes under\--predictions $4\times$ more heavily than over-predictions -- under-predictions cause power budget violations in our optimization problem. The input feature vector consists of the power mode configuration: 
$[cores,cpuf,gpuf,memf]$.

We include the inference $bs$ as an extra feature for inferencing.
 All input features are normalized using \textit{StandardScaler}.
The output layer, in both cases, returns the predicted minibatch time and power.

The NN model is trained for 1000 epochs on different power mode samples, profiled \textit{a priori}, as configured later.
We use an 80:20 train:test split and enable model checkpointing to retain the best model weights from training. After this, a Pareto is constructed from the NN's predictions and is looked up to find the best \textit{estimated} solution.

\textit{Limitations:~} There are two main limitations of NN based time and power predictions for our workload. Firstly, NN requires a lot of profiling samples, typically 100+ power modes, to achieve a good MAPE when power modes are sampled randomly. Secondly, a low MAPE does not always lead to good optimization solution, as shown later in the Results (\S\ref{NNresult}). This is because MAPE is a global metric that reflects average error across \textit{all} power modes (100s). However, for the optimization problem, accuracy of the power modes (10s) lying on the power--time \textit{Pareto} is more important. Having high errors over a few of these Pareto power modes can lead to high optimization errors even if the overall prediction MAPEs are low. 

We address these limitations in our ALS approach.

\subsection{Active Learning-based Sampling (ALS)}
Active Learning (AL)~\cite{activelearning} is an ML technique that minimizes the need for large amounts of labeled data, which can be costly to acquire. 
It picks the most beneficial points to label from the sample space instead of random sampling.
Here, we use AL to select the most beneficial power modes to profile with the goal of reducing the modes profiled.

\subsubsection{ALS Intuition} 
There are several regression-based AL methods. We extend Greedy Sampling on the Output (GSy)~\cite{ALR-greedy} that builds an initial NN model using a few random samples in the input space. The NN model used is the same as described above in \S~\ref{sec:nn}. Using predictions from this model, future samples are picked iteratively to have diversity in the output space -- points predicted to be farthest from prior samples. 

The key intuition is that we want to increase the diversity of power and time values for the selected power modes. This means that the sampling of values from the input space (CPU cores, frequencies, etc.) can be non-uniform depending on their impact on the output space. E.g., we may select nearby values in the input dimension to sample if small changes in input values for that region cause larger changes in the output values. This targeted profiling helps ALS generalize to a variety of configurations using minimal profiling and differentiates it from baselines. For instance, for a training workload, the goal is to build a representative set of power mode samples that can be used to provide solutions to a wide variety of problem configurations such as power budgets. Therefore, in each step, using the predictions of the NN model, power modes that have under-represented output power loads in the existing profiled samples are chosen to profile next.

While targeted sampling improves the quality of samples and reduces profiling overhead, it still does not improve the optimization performance. As described earlier, the NN learns to predict Pareto points accurately by being trained on them. But its mispredictions on non-Pareto points can lead us to mistakenly think that the power mode lies on the Pareto (false positives), leading to suboptimal power mode selection and poor performance. To overcome this limitation, we make a small but crucial change: \textit{we use the NN model in guiding the power mode selection for profiling, but do not use its predictions directly in the optimization process.} The profiling data from the sampled power modes is directly used to construct a partial \textit{observed} Pareto, which is then looked up to find a solution to the optimization problem. Thus, there are no prediction errors with ALS because there are no predictions involved in the optimization process.

\subsubsection{ALS for Standalone Training} 
We adapt GSy to our problem by first training a NN model (\S~\ref{sec:nn}) on the profiling data of $10$ random power modes (\circled{1} in Figure~\ref{fig:ALStrain}, L~\ref{alg2:train:start}--\ref{alg2:train:end} of Algorithm~\ref{algo:als}). 
This NN model predicts the time taken and power load for training/inferencing the target DNN model given a power mode as the input feature.
E.g., for \resnet training on Orin AGX, the initial $10$ random power modes we sample return powers ranging from $14.7$W to $49.3$W.
Particularly, the power values are $14.7, 15.8, 16.6, 19.7, 21.6,$ $22.8,$ $28.0, 28.1, 38.0$ and $49.3$W, which are not necessarily uniformly spaced even though the input power modes were uniformly sampled. These power modes and power values are used to train a NN model to predict power usage.
The profiling similarly returns values for training time, and we train another NN model to predict training time given the power mode.

Based on the predictions of the trained NN models for time and power (\circled{2}), we build a \textit{predicted Pareto front~\footnote{Here, the Pareto front is a set of trade-off points that have the least Y-axis value (time) for any X-axis value (power), or vice versa.}} of training time versus power. We retain only the power modes that lie on the predicted Pareto since the goal is to minimize training time (\circled{3}, L~\ref{alg2:pred:start}--\ref{alg2:pred:end}). E.g., from the NN models trained using the first set of power modes, $31$ power modes fall on the predicted Pareto.

As the NN is trained on incrementally more profiling data from the Pareto region, it gets better at predicting time and power for the Pareto optimal power modes and therefore in guiding power mode sampling.

We then greedily sample $5$ more power modes based on their power load predicted by the NN model such that the modes are \textit{farthest away} from previously profiled modes in terms of \textit{predicted power}, to
maximize the power diversity (\circled{4}, L~\ref{alg2:greedy:start}--\ref{alg2:greedy:end}). We choose power diversity and not time because the Pareto has a steep decline in time with a small change in power for lower power ranges. Thus, sampling based on time diversity would lead to an inefficient oversampling for lower power ranges, and undersampling in the higher power ranges.

E.g., in the earlier example, we select $4$ more power modes whose power loads predicted by the NN model are well spaced out at $25$W, $31.5$W, $32.5$W and $46.4$W. This helps add to the power diversity of the initial $10$ as there were no power modes between $22.8-28$W, between $28.09-38$W and between $38-49.33$W

We profile these $5$ modes \circled{5}, and use their observed power and time to retrain the NN using these additional samples \circled{6} (L~\ref{alg2:train:start}--\ref{alg2:train:end}). We repeat this for $8$ rounds, adding $5$ new power modes in each, and use these $50$ sampled power modes \circled{7} to train increasingly more accurate NN models to predict power and time. We also use these profiled time and power values to incrementally build better observed Paretos \circled{8}. Increasing the output power diversity helps solve a wide range of power constraints in the problem space. The number of initial samples and samples per round are hyper-parameters.

Here, the trained NN model only guides ALS in sampling points to construct the partial Pareto front. 
While the model accurately predicts the time and power for Pareto points, its mispredictions for non-Pareto points  results in sub-optimal solutions, preventing its direct use. We can then solve any problem that uses the same DNN training model, looking it up in the partial Pareto front and returning the optimal power mode within the given power budget \circled{9}.

\subsubsection{Changes for Standalone Inference} Unlike training, where power is the only constraint, inference also has to meet latency constraints. To build a representative set of power modes and $bs$ that generalize to a variety of latency limits and arrival rates, we perform ALS over $4$ quadrants. 

\begin{figure*}[h]
\vspace{-0.1in}
    \centering
    \subfloat[ALS for Standalone Inference]{%
        \includegraphics[width=0.80\textwidth]{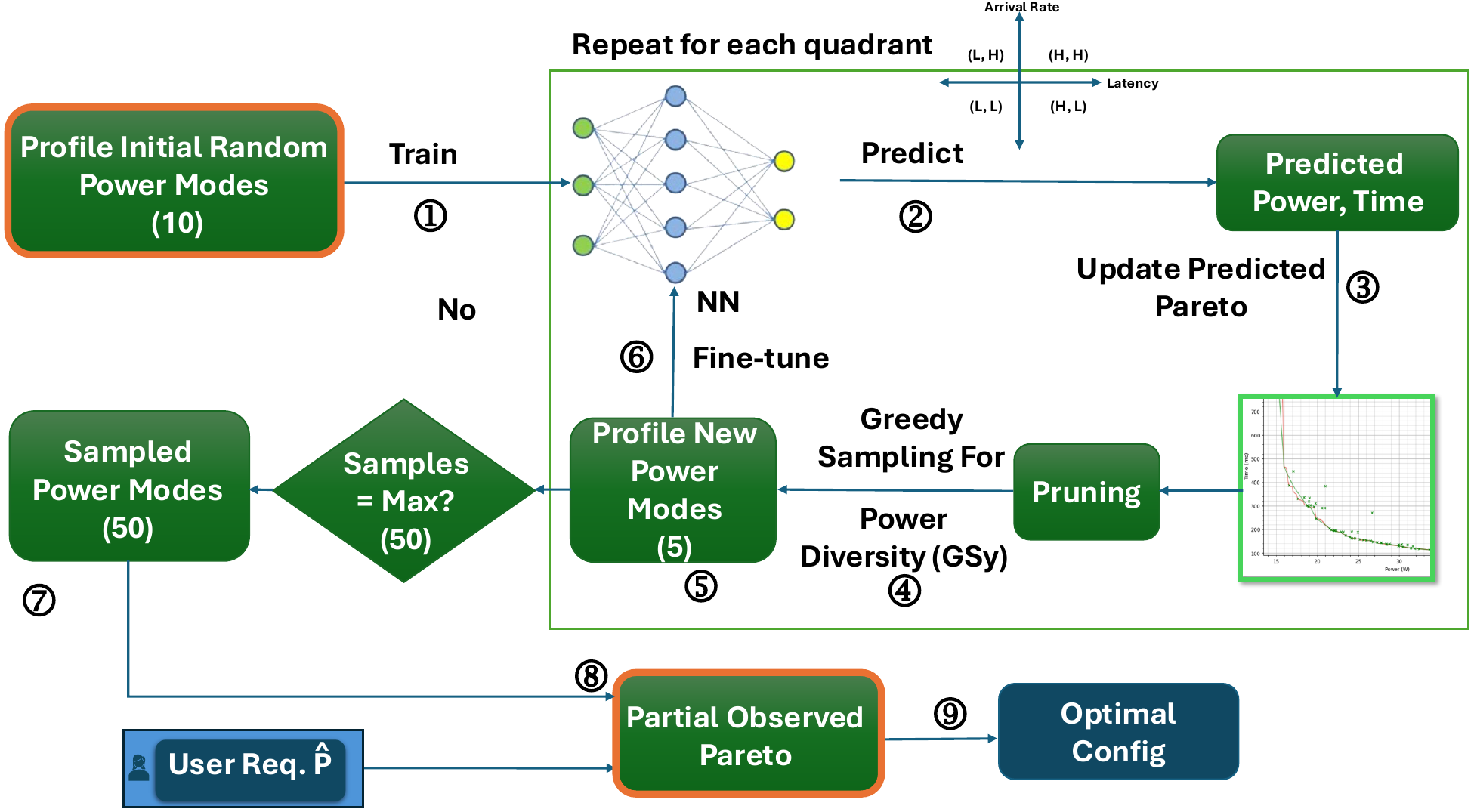}%
        \label{fig:ALSinfer}
    }\\
    \subfloat[GMD for Standalone Inference]{%
        \includegraphics[width=0.55\textwidth]{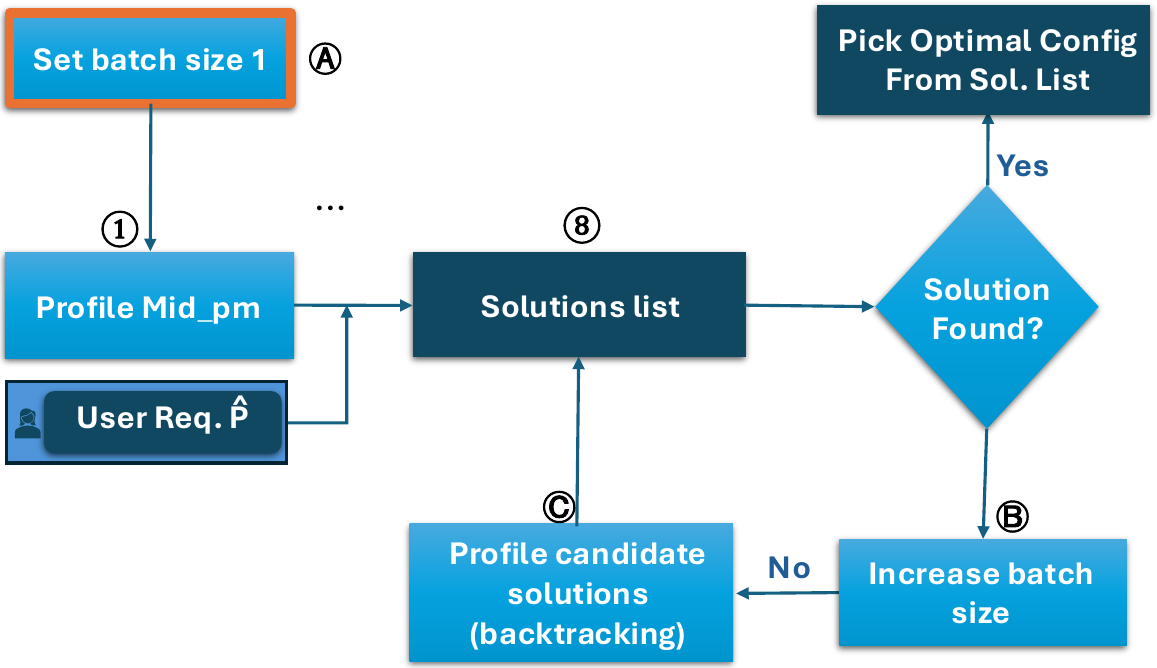}%
        \label{fig:GMDinfer}
    }\\
    \subfloat[GMD for Concurrent Train. and Inf.]{%
        \includegraphics[width=0.55\textwidth]{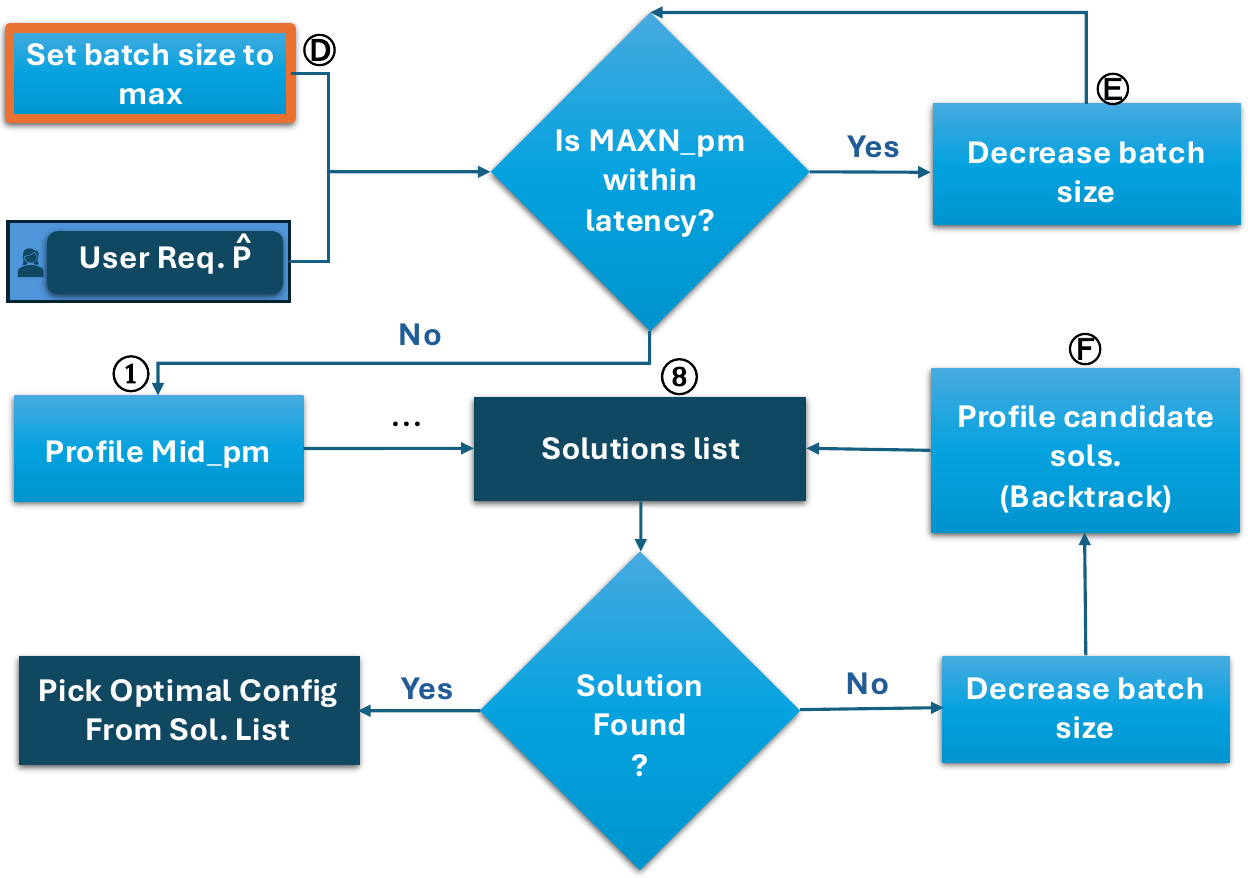}%
        \label{fig:GMDconcur}
    }
    \caption{GMD and ALS for Inference and Concurrent Workloads}
    \label{fig:flowcharts}
\end{figure*}

As shown in Figure~\ref{fig:ALSinfer}, we divide the range of possible latency and arrival rates into $2$ equal regions each (low, high). This results in $4$ possible quadrants. We initialize the ALS model with $25$ power modes ($5$ for each $bs$) and visit each quadrant, round-robin. In a quadrant, we prune out predictions that do not meet the arrival rate/latency ranges for that quadrant. We prune conservatively, removing only modes that do not meet the peak latency or the lowest arrival rate. From modes that remain, we perform the same steps as in standalone training: construct a predicted Pareto, sample the top-$5$ modes based on the predicted power load, profile them and fine-tune the NN model. This repeats for all $4$ quadrants, giving $20$ points per round. We perform $6$ such rounds, with up to \textit{$145$ points}:  initial $25$ points$+6$ rounds$\times 20$ points; it can be fewer if there are fewer points on the Pareto fronts of the initial rounds. 
This Pareto front can solve problems for the same inferencing DNN model but with varying arrival rates, latency and power budgets.

\subsubsection{Changes for Concurrent Workload} We use a quadrant-based approach here as well. The key distinction is that after pruning, we perform a Pareto of the Predicted training throughput vs. Predicted power 
(dominant power among training and inferencing) since the primary goal here is to maximize the throughput. The other steps are the same: retain the Pareto points and pick the top $10$ in every quadrant based on power sampling. This repeats for all $4$ quadrants over $3$ iterations, again resulting in $\leq 145$ points.

\subsection{Extension to Dynamic Rates and Concurrent Inference}
\textit{Dynamic Rates:} GMD and ALS can be extended to support inference workloads with dynamic arrivals too. Since ALS can generalize to a range of arrival rates, the profiled samples can be used as is for inference workloads with dynamic rates. If the rate goes well beyond the range on which ALS was run, it can incrementally run to sample a few more points on the additional range. For GMD, the only change is that we leverage historical profiling data for a workload whenever available and perform additional profiling only when the existing optimal solutions do not satisfy latency. Since GMD navigates the search space based on power, it does not need to be re-run from scratch when the arrival rate changes. Instead, if existing solutions violate latency for a new arrival rate, it quickly backtracks to a higher batch size to find a solution. Thus, both methods can extend easily and performantly to support dynamic rates.

\textit{Concurrent Inference:} While interactive inference workloads such as autonomous navigation~\cite{vehiclecontrol} and augmented reality~\cite{AR_edge} are latency sensitive and have a tight deadline, there are also non-interactive workloads such as offline video analysis~\cite{zhang2015video}, predictive maintenance in Industry 4.0~\cite{predictivemaintenance} that are more focused on throughput improvement by batch processing compute-intensive tasks. Therefore, similar to prior works~\cite{kalmia_infocom,azure_trace}, we extend GMD and ALS to support two concurrent inference workloads of these two types: an \textit{urgent} inference workload with a latency deadline, and a \textit{non-urgent} inference workload with throughput as the QoS. Our goal is to maximize non-urgent inference throughput without violating the deadline of the urgent inference while staying within the power budget. This is conceptually similar to concurrent training and inference, and therefore both GMD and ALS can be used as is.

\subsection{Summary of Strategies}
Unlike GMD which solves for a specific problem configuration of the given workload, ALS can generalize for all problem configurations (power/latency budget, arrival rate) of the workload using the same sampled power modes.
If the workload changes or if run on a new type of edge device, then both strategies will need to be re-run using the newly profiled data for the device/workload.

\section{Experimental Setup}
\label{sec:setup}

\begin{figure}[t]
\vspace{-0.1in}
\centering
\captionof{table}{NVIDIA Jetson Orin AGX Specifications}
\subfloat[Hardware Specifications]{%
\label{tbl:jetsonspecs}
\small
\setlength{\tabcolsep}{1.5pt} 
\renewcommand{\arraystretch}{0.9}
\centering
\begin{tabular}{l|r}
\hline
\textbf{Feature} & \textbf{Orin AGX} \\
\hline\hline
CPU Architecture & ARM A78AE \\ 
\noalign{\global\arrayrulewidth=0.1pt}\arrayrulecolor{lightgray}\hline
\noalign{\global\arrayrulewidth=0.4pt}\arrayrulecolor{black}
CPU Cores & 12 \\
\noalign{\global\arrayrulewidth=0.1pt}\arrayrulecolor{lightgray}\hline
\noalign{\global\arrayrulewidth=0.4pt}\arrayrulecolor{black}
GPU Architecture & Ampere \\
\noalign{\global\arrayrulewidth=0.1pt}\arrayrulecolor{lightgray}\hline
\noalign{\global\arrayrulewidth=0.4pt}\arrayrulecolor{black}
CUDA,Tensor Cores & 2048,64 \\
\noalign{\global\arrayrulewidth=0.1pt}\arrayrulecolor{lightgray}\hline
\noalign{\global\arrayrulewidth=0.4pt}\arrayrulecolor{black}
RAM (GB), Type & $32$, LPDDR5 \\ \hline
Max Power (W) & 60 \\ \hline
Form factor (mm) & $110$$\times$$110$$\times$$72$ \\ \hline
Price (USD) & \$1999 \\ \hline
Accelerators & DLA, PVA \\ \hline
\end{tabular}
}
\subfloat[Power Modes. Freq. in MHz]{%
\label{tbl:jetsonpowerspecs}
\small
\begin{tabular}{l|r}
\hline
\textbf{Features} & \textbf{Orin AGX} \\
\hline\hline
CPU cores & 1 to 12 \\ 
\hline
\# CPU freqs. & 29 \\ 
\hline
CPU Freqs. & 115--2200 \\ 
\hline
\# GPU freqs. & 13 \\ 
\hline
GPU Freqs. & 115--1300 \\ 
\hline
\# Mem freqs & 4 \\ 
\hline
Mem. Freqs. & 665--3200 \\  
\hline
\# Power modes & 18,096 \\
\hline
\end{tabular}
}
\\
\subfloat[Power modes used in experiments]
{\label{tbl:powermodes_exp}
\small
\setlength{\tabcolsep}{1.5pt} 
\begin{tabular}{l|r|r}

\hline
\textbf{Features} & \textbf{Range} & \textbf{\# of values}\\
\hline\hline
CPU cores & 4--12 & 3 \\ 
\hline
CPU Freqs. & 422--2200 & 7 \\ 
\hline
GPU Freqs. & 115--1300 & 7 \\ 
\hline
Mem. Freqs. & 665--3200 & 3 \\  
\hline
\multicolumn{2}{l|}{\# Power modes} & 441 \\ \hline
\end{tabular}
}
\end{figure}

\paragraph{Hardware and Configuration}
We use the NVIDIA Jetson Orin AGX developer kit~\cite{Orin}, whose specifications are given in Table~\ref{tbl:jetsonspecs} and~\ref{tbl:jetsonpowerspecs}. It supports over $\approx 18,000$ power modes. We set the fan speed to max to avoid any thermal throttling. We also disable DVFS so that the configured frequencies remain static during our experiments. 
We do not use or alter custom accelerators, PVA, and DLA.
We use the onboard INA3221 power sensor to get power measurements.

\begin{table*}[t]
\vspace{-0.1in}
\setlength{\tabcolsep}{1pt}
\footnotesize
\caption{DNN Models and Datasets used in Experiments. All models are trained with $bs=16$.}
\vspace{-0.1in}
\label{tbl:modeldataset}
\begin{tabular}
{L{1.2cm}||p{1.3cm}|p{2cm}|p{1cm}p{1cm}p{1cm}||L{2cm}|p{1cm}p{0.8cm}}
\hline
  \bf{Type} & \bf{Task} & \bf{Model} & \bf{\# Layers} & \bf{\# Params} & \bf{FLOPs}$^\dagger$ & \bf{Dataset} & \bf{\# Samples} & \bf{Size} \\
  \hline\hline
 Training, Inference& Image classif. & \textbf{{MobileNet}-v3~\cite{mobilenet}}& 20  & $5.5M$ & $225.4M$ & \textbf{GLD23k~\cite{tensorflow_gld23k}} & 
 $23k$  
 & $2.8GB$ \\
  \hline
 Training & Image classif. & \textbf{\resnet-18~\cite{resnet}}& 18  & $11.7M$    
  & $1.8G$ &  \textbf{ImageNet Val.~\cite{imagenet}} & 
 $50k$ 
 & $6.7GB$\\
  \hline
 Training, Inference & Object detection &\textbf{\yolo-v8n~\cite{yolo}} & 53 &  $3.2M$      &  $8.7G$ &  \textbf{COCO minitrain~\cite{coco_minitrain}} &  
 $25k$
 & $3.9GB$\\
  \hline
 Training & Question ans. & \textbf{BERT base~\cite{bert}}& 12 & $110M$  & $11.5T$ 
 & \textbf{SQuAD V2.0 Train~\cite{squad}} & 
 $130k$ 
 & $40.2MB$ \\
    \hline
 Training, Inference & Next word pred. & \textbf{LSTM~\cite{LSTM}}& 2  & $8.6M$  & $3.9G$ & \textbf{Wikitext~\cite{wikitext}} & 
 $36k$ 
 & $17.8MB$ \\
\midrule
\midrule
 Inference & Image classif. & \textbf{\resnet-50~\cite{resnet}}& 50  & $25.6M$ &  $3.8G$ 
 & \textbf{ImageNet Val.~\cite{imagenet}} & 
 $50k$ 
 & $6.7GB$ \\
  \hline
 Inference & Question ans. & \textbf{BERT Large~\cite{bert_large}}& 24 & $340M$  & $43.7T$
 & \textbf{SQuAD V2.0 Dev~\cite{squad}} & 
 $12k$ 
 & $4.2MB$ \\
\hline

\multicolumn{8}{p{11cm}}{$^\dagger$~\textit{As per the typical practice, FLOPs reported correspond to a forward pass with minibatch size 1.}}
\end{tabular}
\end{table*}

\paragraph{Training and Inference workloads}
We choose five popular DNN workloads each for DNN training and inference (Table~\ref{tbl:modeldataset}). Two are image classification tasks, one does object detection, one is for question answering, and the last for next word prediction. These are representative of typical use cases and workloads on edge devices~\cite{AbelmoniemREFL, jallepalli_federated_yolo, tian_fedbert}. We pick a variety of model architectures (CNN, RNN and transformer) and tasks with sufficient diversity in dataset sizes ($17.8$MB--$6.7$GB), model sizes ($3.2$M--$110$M parameters) and computational requirements ($225$M--$11.5$T FLOPS). The power budgets, arrival rates and latency budgets used for standalone inference are reported in Section \S\ref{sec:infresults} and for concurrent inference in Section \S\ref{sec:concresults}

\paragraph{Software and Libraries}
The Orin AGX runs Ubuntu 20.04 LTS with kernel version 5.10.65. We use NVIDIA JetPack v5.0.1 with CUDA v11.4. We configure this with PyTorch v1.12 and torchvision v0.13. The PyTorch's Dataloader runs with $4$ worker processes wherever possible~\footnote{Due to a bug in YOLO v8 we run it with only $1$ worker. This is fixed in a later PyTorch version, but it is not compatible with our JetPack version (\url{https://github.com/pytorch/pytorch/issues/48709})}. We use a minibatch size of $16$ for all training workloads, and run $5$ minibatch sizes: $(1,4,16,\-32,64)$ for inference workloads~\footnote{We do not run BERT inference for minibatch size $64$ as it takes $>20$s per minibatch at lower power modes.}. 

\paragraph{Profiling Setup and Metrics}
We sample the power load every 1s using \texttt{jtop}, which is a wrapper around NVIDIA's \texttt{tegra\-stats} library. This gives the Jetson module's power and omits the power consumption of peripherals (USB ports etc.), which is negligible~\cite{prashanthi2023sigmetrics}. We instrument PyTorch to measure the minibatch execution time using \texttt{torch.cuda.event} with the \texttt{synchronize} flag to accurately measure GPU time. We confirm that our profiling overhead is minimal and does not affect the workload performance. 

When we profile a power mode, we run the DNN (training or inference) for $\approx 40$ minibatches and record the minibatch time and sampled power. The first minibatch takes long to execute, possibly due to PyTorch's profiler picking the best implementation, and we hence discard this sample. The power also takes $2$--$3$s to stabilize. 
We detect this stabilization point and only use profiling data beyond this.

\paragraph{Data Collection for Ground-Truth and ``Optimal'' Solution}
We pick $441$ power modes that are uniformly distributed through the $\approx 18,000$ power modes. This spans 3 core counts, 7 CPU frequencies, 7 GPU frequencies, and 3 memory frequencies (Table~\ref{tbl:powermodes_exp}). We run our workloads using all these $441$ power modes.
We determine the nominal ``optimal'' solution for a problem from within this large (but not exhaustive) ground truth space.
We also perform a smaller set of runs for interleaving and confirm that interleaved minibatch times match the sum of the minibatch times for training and inference. Also, the interleaved power equals the maximum of the training and inference power. 

\paragraph{Baseline Strategies}
\textbf{(i)~Random.} We use 2 static/random sampling baselines. For the training workloads, these are \textit{RND50} and \textit{RND250}, which use $50$ and $250$ modes out of the $441$. For inferencing, we use \textit{RND150} and \textit{RND250}, where RND150 picks $30$ power modes at random and profiles each for all $5$ minibatch sizes; RND250 uses $50$ modes. This profiling data is used to construct a partial Pareto front of time and power. We then look up this Pareto to select the best solution for a given problem.

\textbf{(ii)~Neural Network (NN).} We train the NN model with 250 random power modes for both inference and training. A Pareto is constructed from its predictions and is looked up to find the best \textit{estimated} solution. The number of power modes for the baselines is chosen such that one random baseline has the same number of power modes as ALS and much higher than GMD e.g., RND50, ALS50, GMD10 for training, and the other random/NN baseline has a higher number than ALS e.g., RND150/NN150 and ALS50 for training.
\textbf{(iii)~Optimal.} We contrast these strategies against the ground-truth nominally optimal solution. We lookup over the Pareto front constructed from all $441$ power modes and $5$ minibatch sizes each that are profiled to return the ``optimal'' power mode that solves the problem.

\section{Results and Analysis}
\label{sec:results}

We conduct a detailed evaluation of our ALS and GMD strategies across $5$ standalone training and inference DNNs, and $5$ concurrent training and inference DNN pairs. We comparatively evaluate them for a total of $273$k problem configurations across varying power/latency budgets and arrival rates.  
The number of power modes profiled is a hyperparameter that varies for training, inference and concurrent workloads. We indicate this in the strategy name,
e.g., ALS50 means that $50$ modes are profiled by ALS while GMD10 means $10$ modes are used. This is chosen in different ways for GMD and ALS: For ALS, we perform experiments to tune this hyperparameter and set the budget accordingly for training, inference and concurrent. For GMD, it is the number of searches after which the space is exhausted. This is again different for training, inference and concurrent.

\subsection{Standalone Training Workloads}
\begin{figure*}[t]
\vspace{-0.1in}
 \centering
\subfloat[\resnet]{%
    \includegraphics[width=0.33\textwidth]{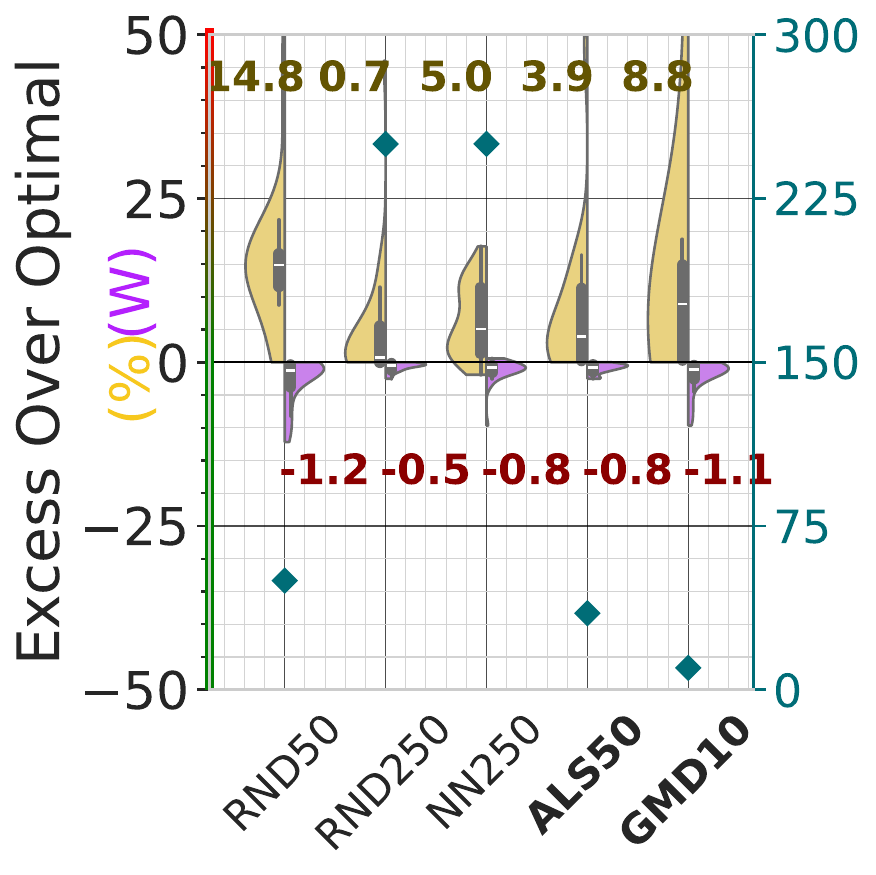}%
    \label{fig:stdtrain_resnet_train}%
  }%
 \subfloat[\mobilenet]{%
    \includegraphics[width=0.33\textwidth]{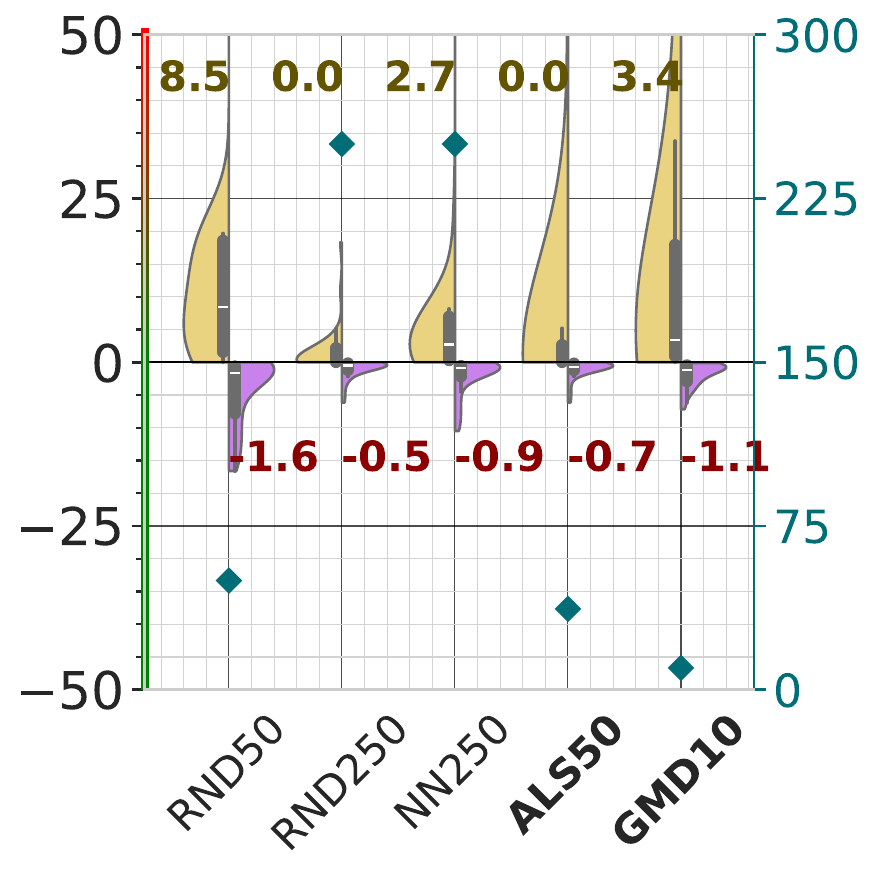}%
    \label{fig:stdtrain_mobnet_train}%
  }%
  \subfloat[\yolo]{%
    \includegraphics[width=0.33\textwidth]{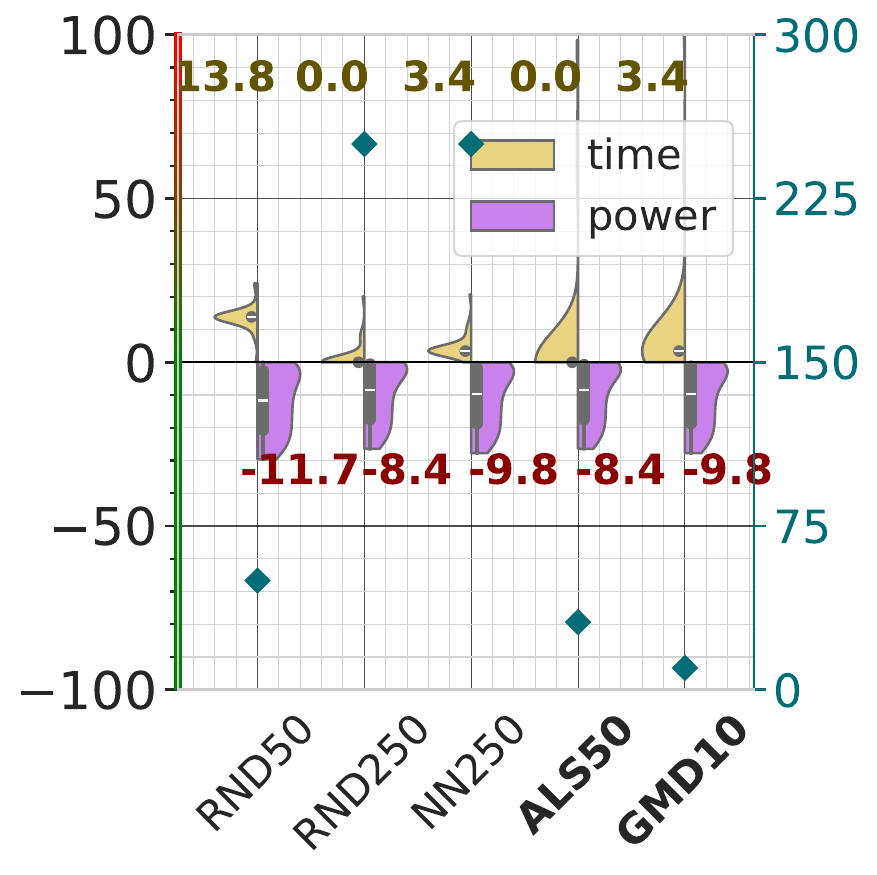}%
    \label{fig:stdtrain_yolo_train}%
  }\\
  \subfloat[\bert]{%
    \includegraphics[width=0.33\textwidth]{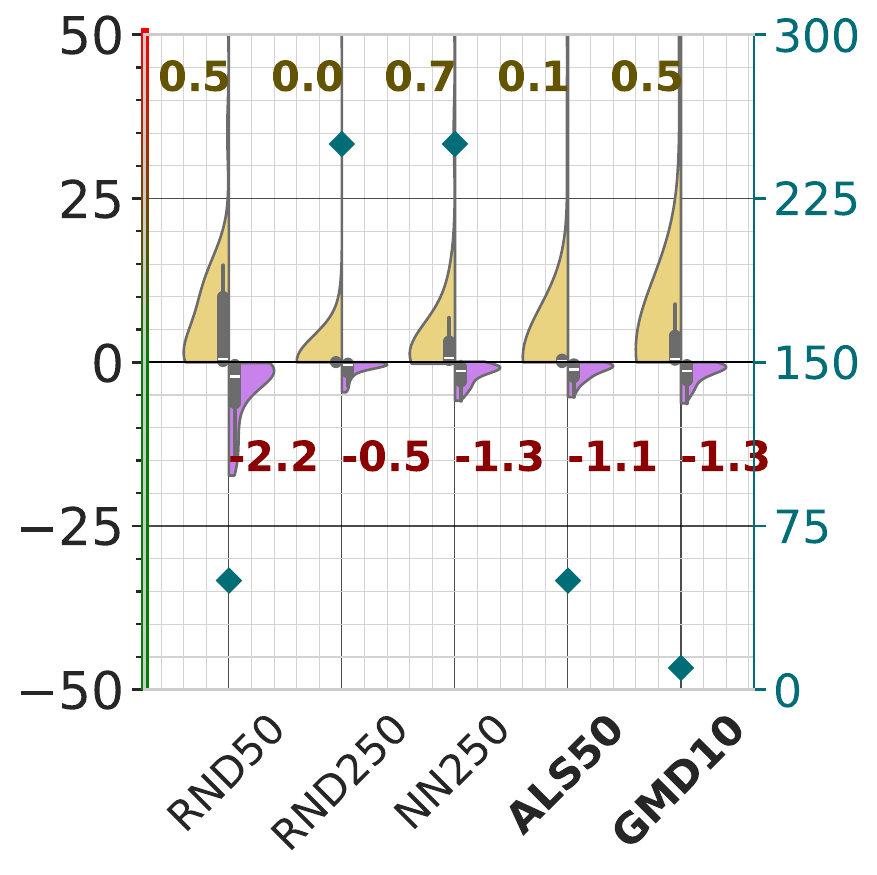}%
    \label{fig:stdtrain_bert_train}%
  }%
  \subfloat[\lstm]{%
    \includegraphics[width=0.33\textwidth]{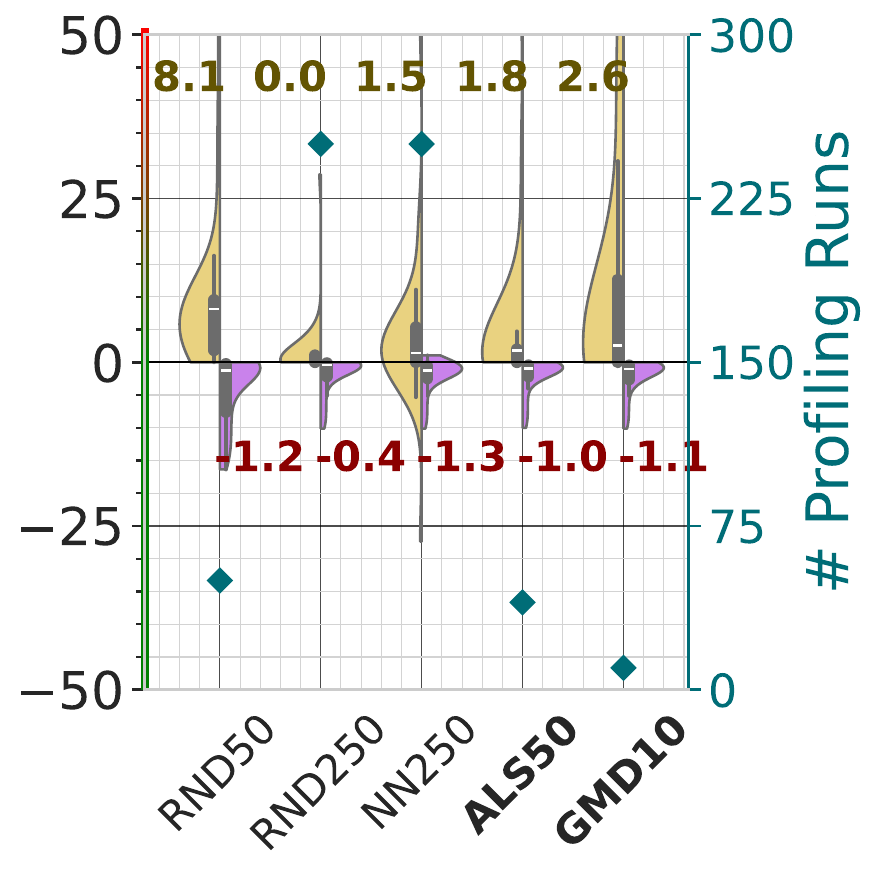}%
    \label{fig:stdtrain_lstm_train}%
  }%
\vspace{-0.1in}
\caption{Excess over Optimal training time (\%) and absolute power (W) for all strategies and baselines on Standalone Training workloads.}

\label{fig:stdtrain_Dual_Violin}
\vspace{-0.15in}
\end{figure*}

We vary the power budget from $10$--$50$W in increments of $1$W to generate problem configurations for all four training DNNs except \bert and $10$--$60$W for \bert since its power load goes up to $60$W unlike the rest. This results in a total of $215$ problem configurations.
We compare the performance of ALS and GMD with the baselines. In Figure~\ref{fig:stdtrain_Dual_Violin}, we report the distance from optimal in the form of time and power violin plots. The medians for time and power are shown as white markers and labeled. Inter-quartile range are shown as black boxes over the violins. Every point in the violin represents a solution (power mode) for a problem configuration.

The \textit{time violin} (yellow, top half in the plot) represents the percentage difference in minibatch time between the strategy and the optimal solution. Since the minibatch size is constant, training throughput is inverse of the training minibatch time; so lower values in the time violin indicate higher throughput.  
A positive value in the time violin indicates that the time taken by a strategy's solution is higher than the optimal, while a negative value means that the strategy picks a power mode that trains faster than the optimal, but with a power violation. A lower time is better, provided there are no power budget violations.

The \textit{power violin} (purple, bottom half) represents the absolute difference in power used by the solution and the given power budget. We also report the number of profiled power modes by each strategy on the right Y axis (teal markers).

\claim{ALS50 outperforms RND50 across all cases even though both profile the same number of power modes}
As seen in Figure~\ref{fig:stdtrain_resnet_train} for \resnet, RND50 exceeds the optimal time by a median of $14.8\%$, whereas ALS is higher than optimal by only $3.9\%$. Likewise, RND50 for \mobilenet exceeds the optimal by $14.8\%$ while the median of ALS exactly matches the optimal. \yolo, \bert and \lstm show median excess times for ALS of $0.0,0.1$ and $1.8\%$, while RND50 has as higher median excess values of $13.8\%, 0.5\%$ and $8.1\%$. For \bert, several solutions lie close to the Pareto for RND50, leading it to approach optimal, though ALS still outperforms it. Additionally, ALS not only has a lower median but also a tighter inter-quartile range  compared to RND50. Hence, ALS returns near optimal solutions to meet the optimization goal of minimizing DNN training time within the power budget, compared to random sampling of the same number of power modes.

\claim{ALS50 often performs as well as RND250, which profiles $5\times$ more power modes}
For \mobilenet and \yolo, we see from Figure~\ref{fig:stdtrain_Dual_Violin} that ALS and RND250 both have a median time that matches the optimal ($0.0\%$). For \bert and \lstm, while RND250 gives optimal solutions, ALS is close at only $0.1\%$ and $1.8\%$ over the optimal time. For \resnet alone, RND250 is better at $0.7\%$ while ALS is at $3.9\%$. So, in $4$ out of $5$ cases, ALS performs close to RND250 that profiles $5\times$ more power modes, showcasing the highly efficient sampling of ALS.

\claim{GMD10 outperforms RND50 even though it profiles $5\times$ fewer power modes}
As seen in Figure~\ref{fig:stdtrain_Dual_Violin}, for \resnet, RND50 has a median excess over optimal of $14.8\%$, whereas GMD has a median of $8.8\%$. Similarly, for \mobilenet, RND50 has a median of $8.5\%$ in comparison to GMD's $3.4\%$, and for \yolo RND50 and GMD have medians of $13.8\%$ and $3.4\%$ respectively. GMD with just $10$ points, outperforms RND50 in all $5$ cases, showcasing that GMD is well-suited to scenarios that need a quick solution as shown in Table~\ref{tbl:PracMat}.

\claim{NN250 is comparable or slightly better than GMD10 while profiling $25\times$ more power modes, but violates power budgets for $2$ out of $5$ DNNs.}
NN is a more sophisticated baseline than RND. For \lstm, NN250 has a median excess time over optimal of $1.5\%$, which is slightly better than GMD's median of $2.6\%$. However, NN violates power budgets by choosing power mode solutions that take lesser training time than the optimal, but consume a higher power that violates the budget -- seen in the negative part in the time violin and the positive part in the power violin. Such violations are seen in $2$ cases (\resnet and \lstm) out of $5$. This is despite NN profiling $250$ power modes compared to $10$ by GMD, a $25\times$ difference, while performing comparably or slightly better in the other $3$ cases.

\claim{NN250 is outperformed by the RND250 random baseline that uses the same number of power modes} \label{NNresult}
In Figure~\ref{fig:stdtrain_Dual_Violin}, RND250 is close to the optimal for \resnet, with an excess time of just $0.7\%$. This is much lower than NN250's median of $5.0\%$. Even for the other $4$ DNN models, RND250 has near optimal median times while NN250 has median excess times over optimal ranging from $0.7$--$3.4\%$.
Even though NN has low MAPE errors of $<3\%$ for time and power predictions across all power modes and DNNs, what matters is the prediction accuracy of non Pareto points for the standalone training workload. The accuracy in this region is not as high, causing mispredictions to be amplified when solving the optimization problem. So, if profiling data for many power modes is available (e.g., $250$ out of $441$ here), it is better to construct and solve using an observed Pareto, which does not suffer from prediction errors, than NN.

\claim{ALS50 and GMD10 use the actual profiling output in decision making, and hence never violate the power budget}
In Figure~\ref{fig:stdtrain_Dual_Violin}, the power violins for ALS and GMD only have negative points, i.e., their solution always stays under the power budget. This is also true for RND. These strategies use the \textit{observed} power values from profiling a power mode and never pick a solution whose ground truth violates the budget. Since NN is a prediction-based method, it can wrongly estimate the time or power, leading to seemingly lower time than optimal but causing a power violation due to prediction errors. This is seen as negative values in time violins and positive values in the power violins for \resnet and \lstm.

\begin{figure*}[t]
\vspace{-0.1in}
 \centering
\subfloat[\resnet]{%
    \includegraphics[width=0.33\textwidth]{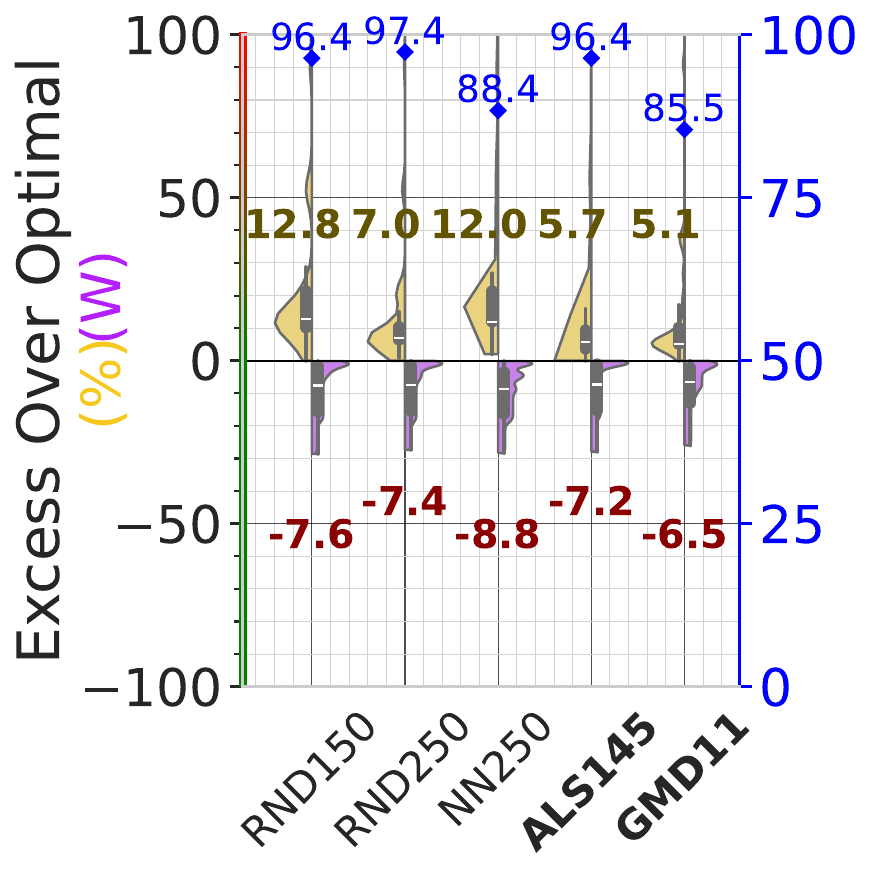}%
    \label{fig:stdinfer_resnet_infer}%
  }%
 \subfloat[\mobilenet]{%
    \includegraphics[width=0.33\textwidth]{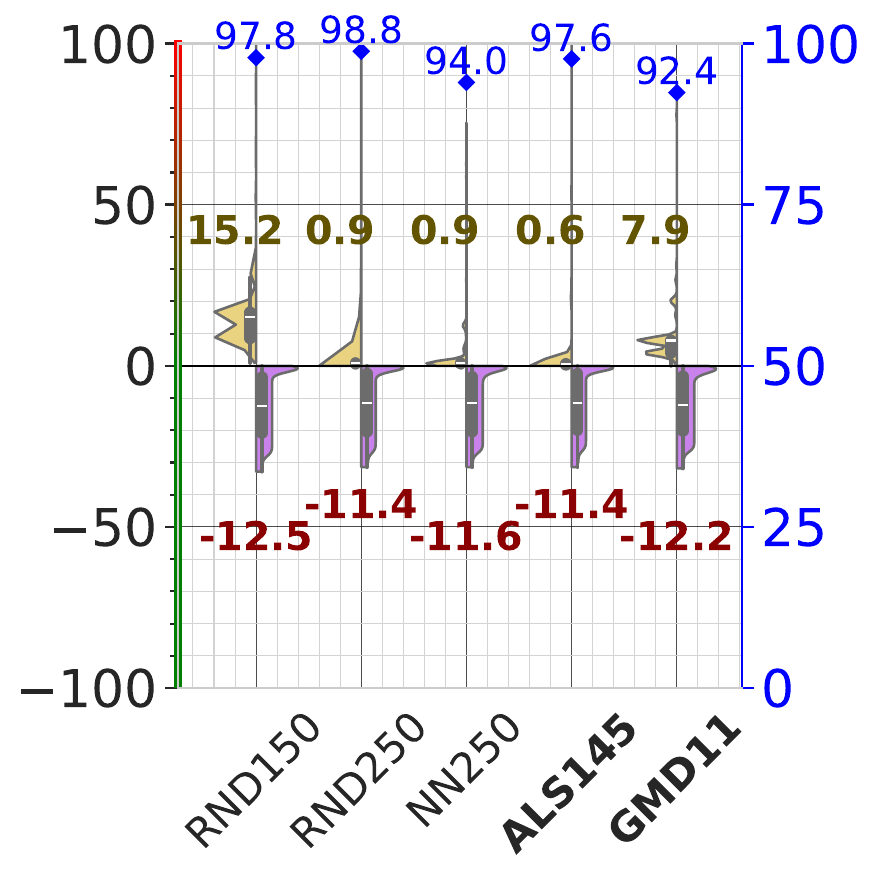}%
    \label{fig:stdinfer_mobnet_infer}%
  }%
  \subfloat[\yolo]{%
    \includegraphics[width=0.33\textwidth]{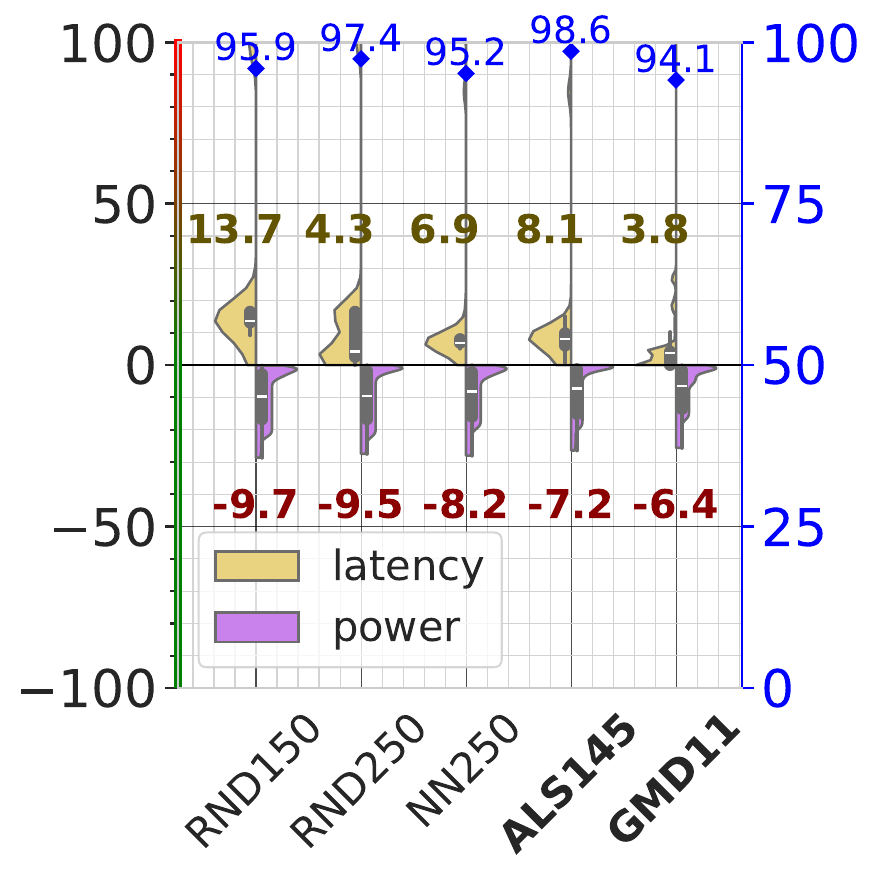}%
    \label{fig:stdinfer_yolo_infer}%
  }\\
  \subfloat[\bert]{%
    \includegraphics[width=0.33\textwidth]{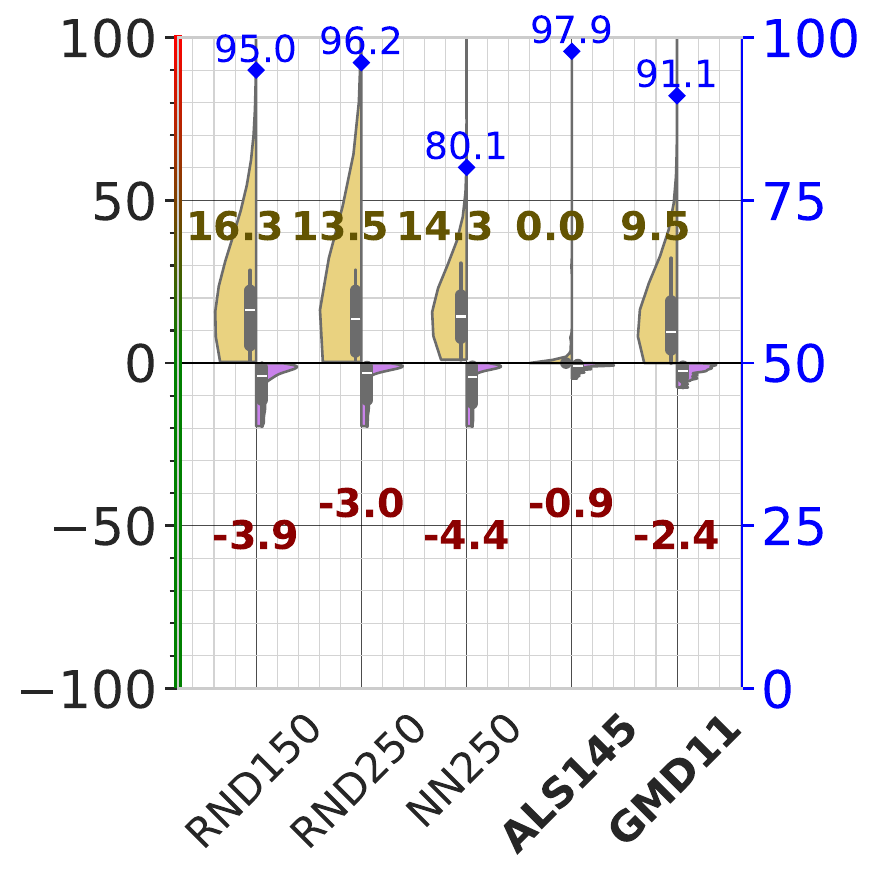}%
    \label{fig:stdinfer_bert_infer}%
  }%
  \subfloat[\lstm]{%
    \includegraphics[width=0.33\textwidth]{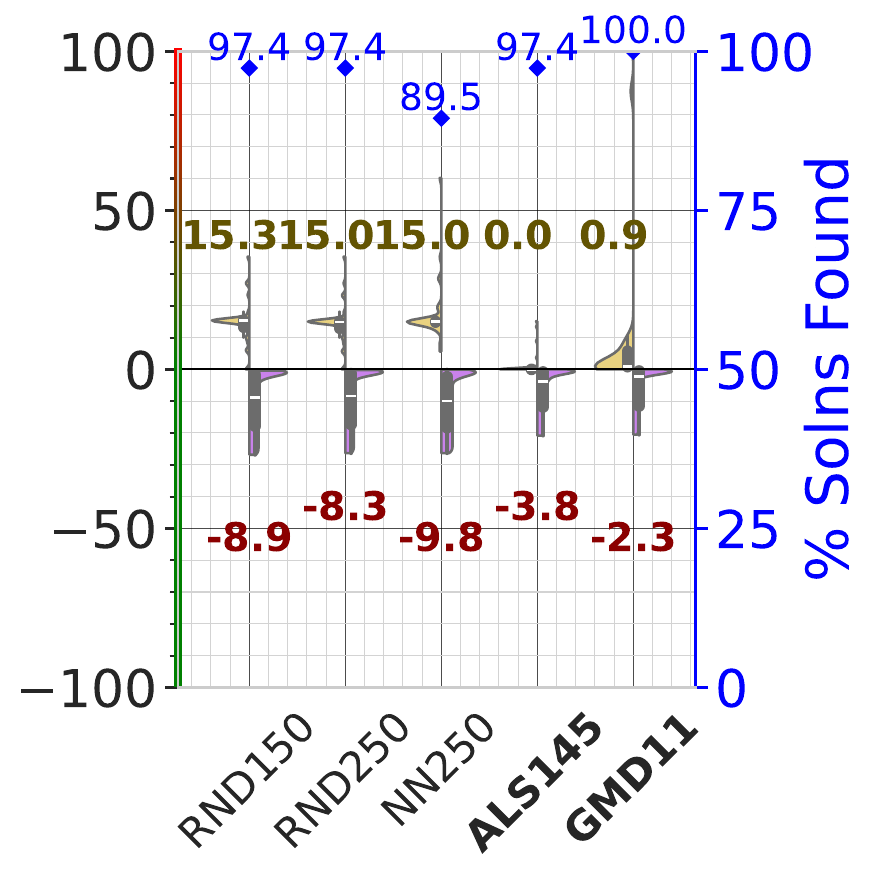}%
    \label{fig:stdinfer_lstm_infer}%
  }%
\vspace{-0.1in}
\caption{Excess over Optimal inference latency (\%) and absolute power (W) for all strategies and baselines on Standalone Inference workloads.}

\label{fig:stdinfer_Dual_Violin}
\vspace{-0.15in}
\end{figure*}

\subsection{Standalone Inference Workloads}
\label{sec:infresults}
Next, we evaluate problem configurations spanning different power budgets, latency budgets and arrival rates for DNN inferencing. For \resnet, \mobilenet, \yolo and \lstm, we vary the power budget from $10$--$50$W in intervals of $1$W, latency budget from $50$--$1000$ms in intervals of $10$ms and arrival rate from $30$--$90$~RPS in steps of $5$. For \bert, which has a much higher inference time, we use a latency budget of $1$--$10$s in steps of $200$ms and an arrival rate of $1$--$5$~RPS in intervals of $1$. We pick these configurations 
based on literature on vision and language models~\cite{clockwork,infaas} and adapt the server latency deadlines to edge accelerator timescales. This gives $\approx 35k$ problem configurations for BERT and $\approx 51k$ each for the other $4$ for a total of $240k$ configurations.
We solve these using the different strategies and report results in Figure~\ref{fig:stdinfer_Dual_Violin}.
Since inference has both latency and power budgets, some strategies are unable to find solutions in some cases even though there is an optimal solution. We report the fraction of problems solved on the right y-axis (blue marker), and unsolved ones are omitted from the violin. When NN's solution violates the latency or power budget, this is also reported as a solution not found.

\claim{ALS145 finds solutions for more than $96\%$ of problems, and outperforms RND150 for all DNN models, with similar profiling overheads} 
Both these strategies perform $145 \sim 150$ profiling runs (power mode+minibatch size).
But RND150 has an excess median latency over the optimal for \bert and \lstm of $16.3\%$ and $15.3\%$ (Figure~\ref{fig:stdinfer_Dual_Violin}), while ALS is significantly better with the median matching the optimal. This holds for \mobilenet, with RND150 and ALS having medians of $15.2\%$ and $0.6\%$. ALS outperforms RND150 for all $5$ DNNs on time. Both find solutions for $>96\%$ of the $35,000+$ configurations.

\claim{ALS145 performs better than RND250 for 4 out of 5 DNNs, despite having a lower profiling overhead}
Increasing the profiling runs for RND to 250 improves its performance to approach ALS145, but with $70\%$ more profiling overheads.
In Figure~\ref{fig:stdinfer_Dual_Violin}, 
for \mobilenet, ALS145 and RND250 are comparable on time, taking $0.6\%$ and $0.9\%$ in excess of optimal time. But ALS starts to perform better for \resnet, at $5.7\%$ which is slightly better than $7.0\%$ for RND250. ALS is clearly ahead of RND250 for \bert and \lstm, where it has optimal median time while we see that RND250 has excess latency of $13.5\%$ and $15.0\%$.
Unlike training, there are $5$ minibatch sizes for inference. So $250$ profiling runs effectively cover $50$ power modes. Yet, RND250 has higher \% excess time compared to RND50 for training. So, $250$ randomly chosen samples are insufficient. ALS exhibits superior sampling quality, achieving better performance with fewer 145 runs.

\claim{GMD11 finds solutions for $>85\%$ of all problems using $\leq 11$ profiling runs and outperforms RND150 for all DNNs}
Unlike ALS, RND and NN, GMD performs profiling on the fly for the specific problem configuration. Yet, it requires $\approx 14\times$ fewer runs and also offers better time performance.
For \resnet, \mobilenet and \yolo, RND150 has median excess time of $12.8\%, 15.2\%$ and $13.7\%$ over optimal, while GMD is much better with medians of $5.1\%, 7.9\%$ and $3.8\%$ (Figure~\ref{fig:stdinfer_Dual_Violin}). GMD outperforms RND150 for the other $2$ DNNs as well.
While RND150 gives solutions for 95\% of problem configurations, GMD falls short due to its budget of only $11$ profiling runs. But it still finds a viable solution in more than $85\%$ of cases.

\claim{GMD11 outperforms NN250 for $4$ out of $5$ DNNs, despite using $\approx 23\times$ fewer profiling runs}
For \resnet, GMD11 has a median of $5.1\%$ excess time over optimal, which is much better than NN's median of $12\%$. Similarly, for \bert and \lstm, GMD has medians of $9.5\%$ and $0.9\%$, which outperform NN's medians of $14.3\%$ and $15.0\%$. For \mobilenet, NN at $0.9\%$ outperforms GMD which is at $7.9\%$. Here, the training data for NN250 are coincidentally close to the Pareto; this is confirmed by the better time even for RND250 ($0.9\%$), which shares the same samples.
In all the other $4$ DNNs, GMD does better, highlighting its ability to find a solution within $11$ runs.

\claim{NN250 performs worse than RND250, and solves fewer problems}
NN continues to perform worse than RND250 for $4$ out of $5$ DNNs, similar to inferencing. E.g., RND250 has a median excess time of $7.0\%$ for \resnet and finds solutions to $97.4\%$ of the problem configurations, while NN has a median of $12.0\%$ and finds solutions to $88.4\%$ of the problems.%

\subsection{Concurrent Training and Inference}
\label{sec:concresults}
\begin{figure*}[t]
\vspace{-0.1in}
 \centering
 \subfloat[\{MNet\ \textit{Train}, \lstm \textit{Infer}\}]{%
    \includegraphics[width=0.3\textwidth]{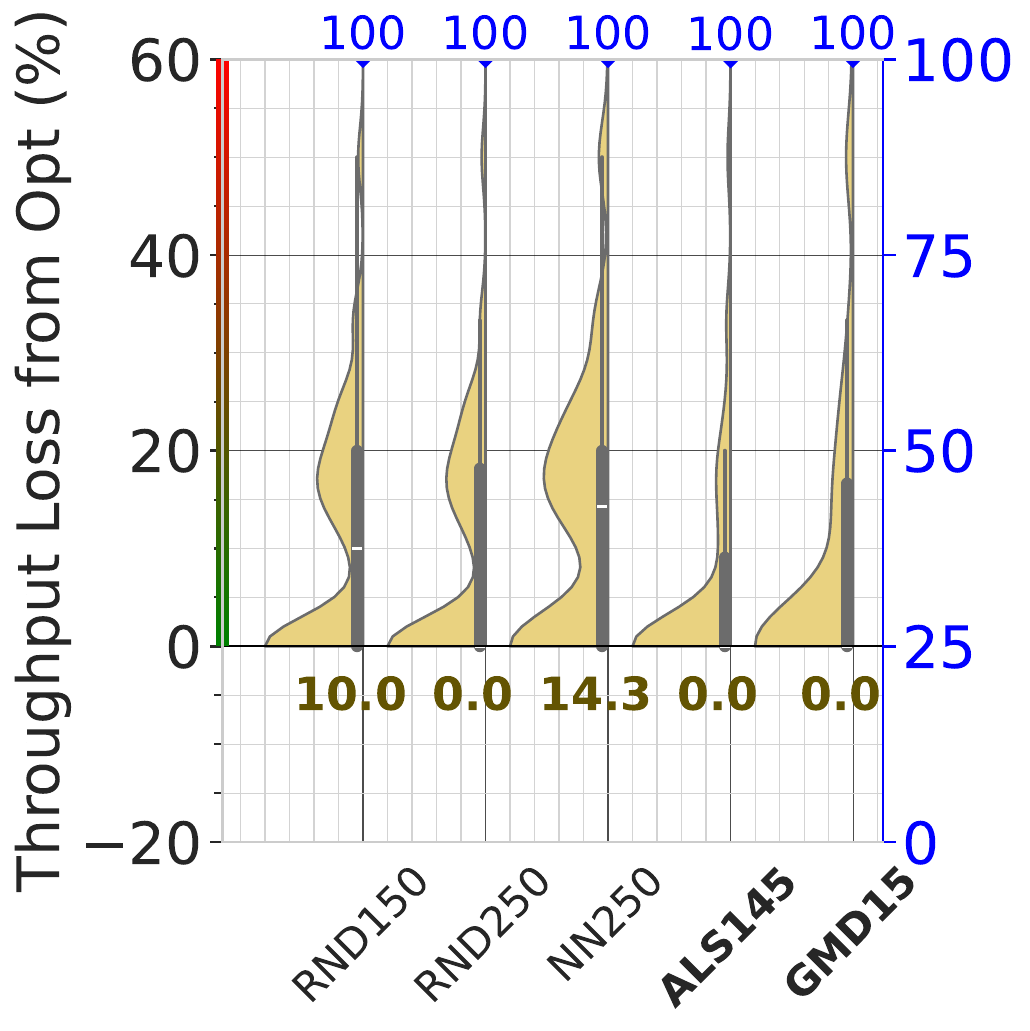}%
\label{fig:interleave_mobnet_train_lstm_infer}%
}%
 \subfloat[\{MNet,MNet\}]{%
    \includegraphics[width=0.3\textwidth]{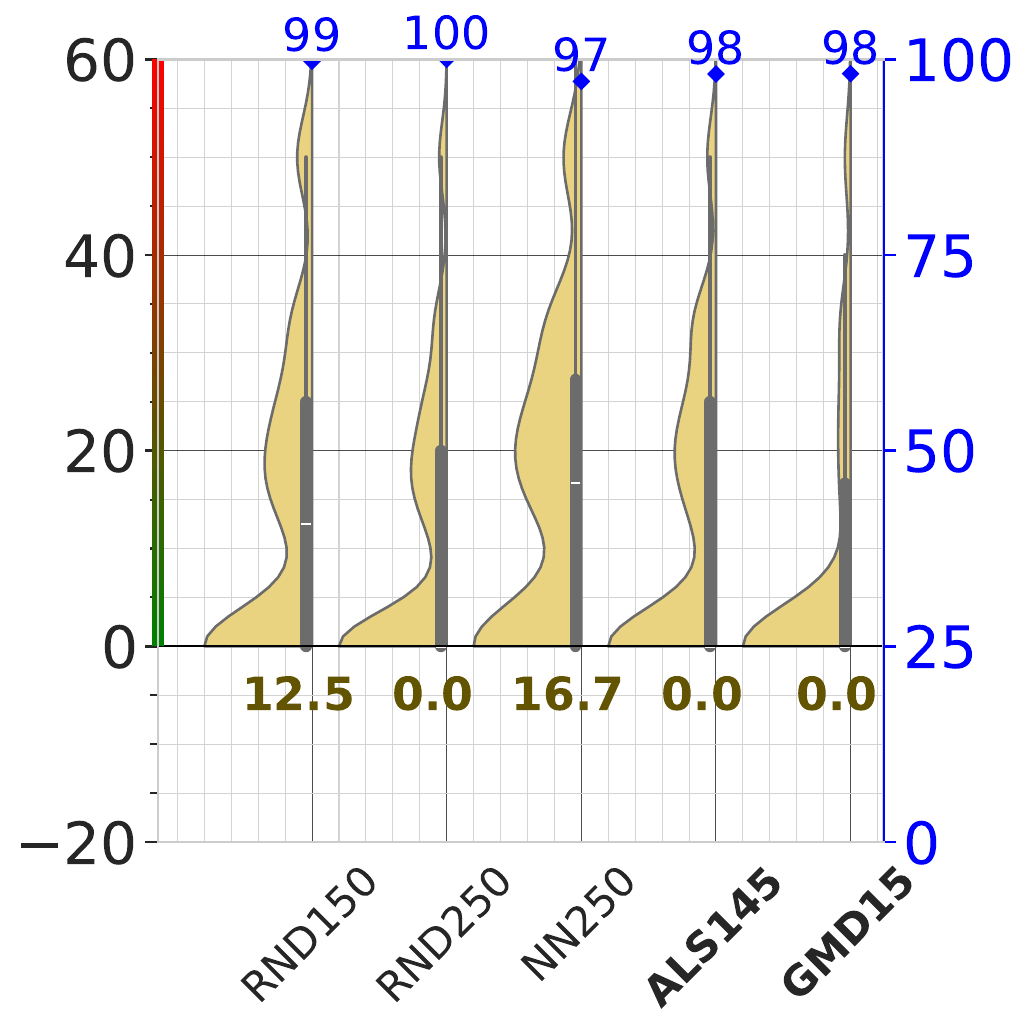}%
\label{fig:interleave_mobnet_train_mobnet_infer}%
}%
 \subfloat[\{\resnet,MNet\}]{%
    \includegraphics[width=0.3\textwidth]{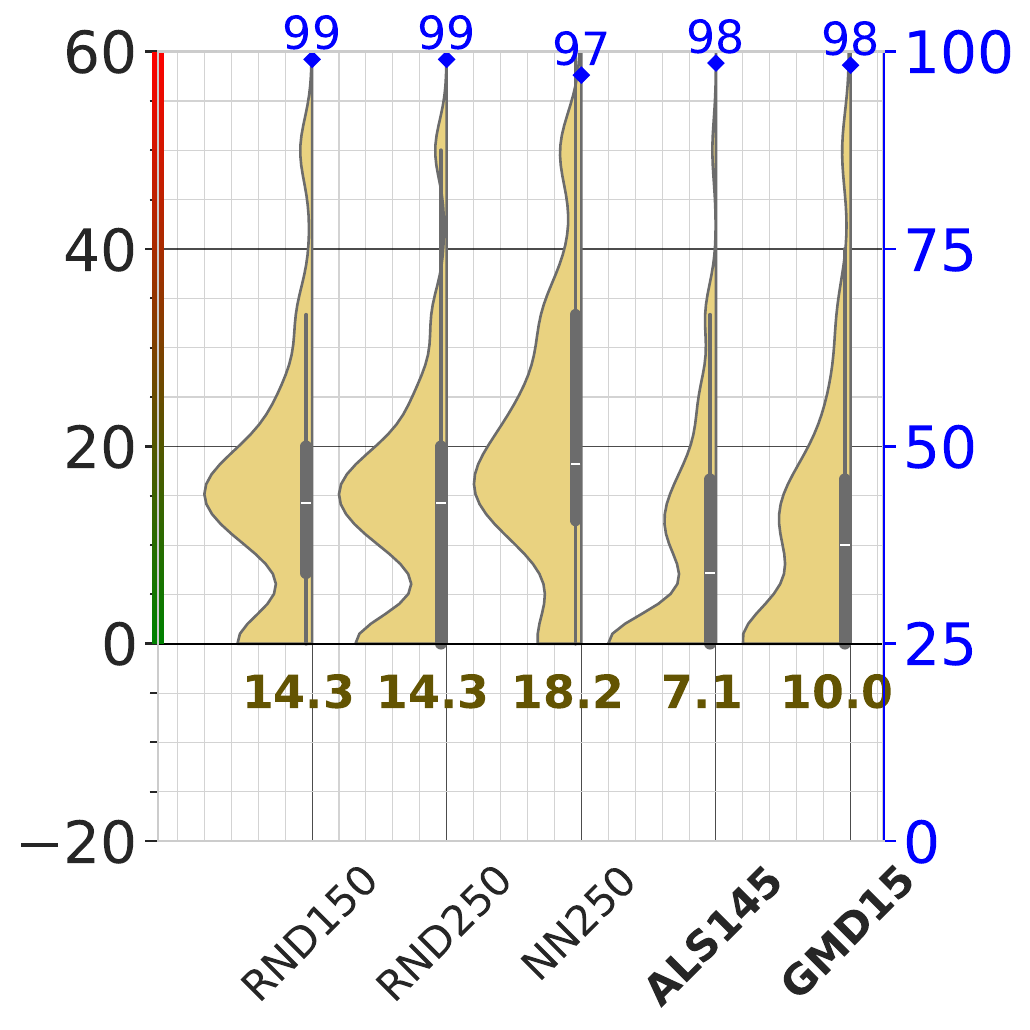}%
\label{fig:interleave_resnet_train_mobnet_infer}%
}\\
 \subfloat[\{\resnet,\bert\}]{%
    \includegraphics[width=0.3\textwidth]{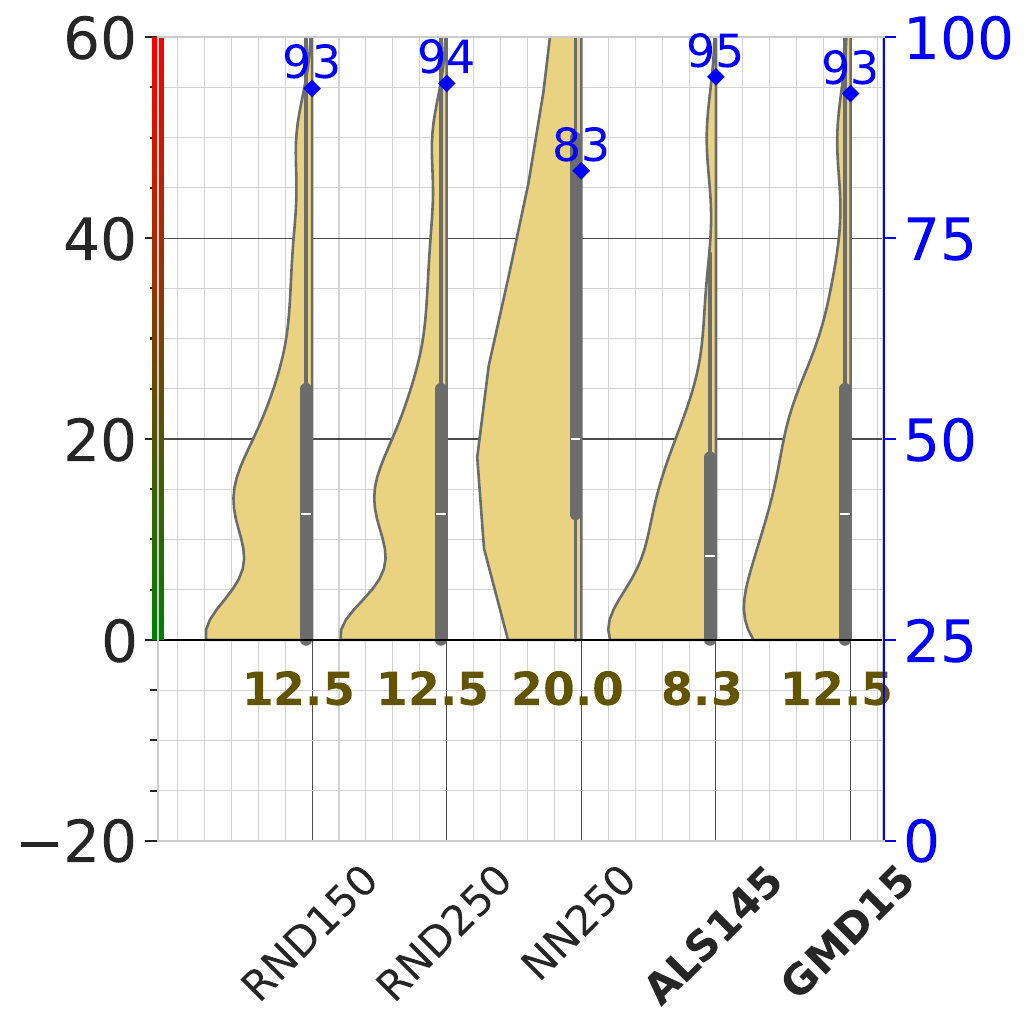}%
\label{fig:interleave_resnet_train_bert_infer}%
}%
 \subfloat[\{\yolo,\resnet\}]{%
    \includegraphics[width=0.3\textwidth]{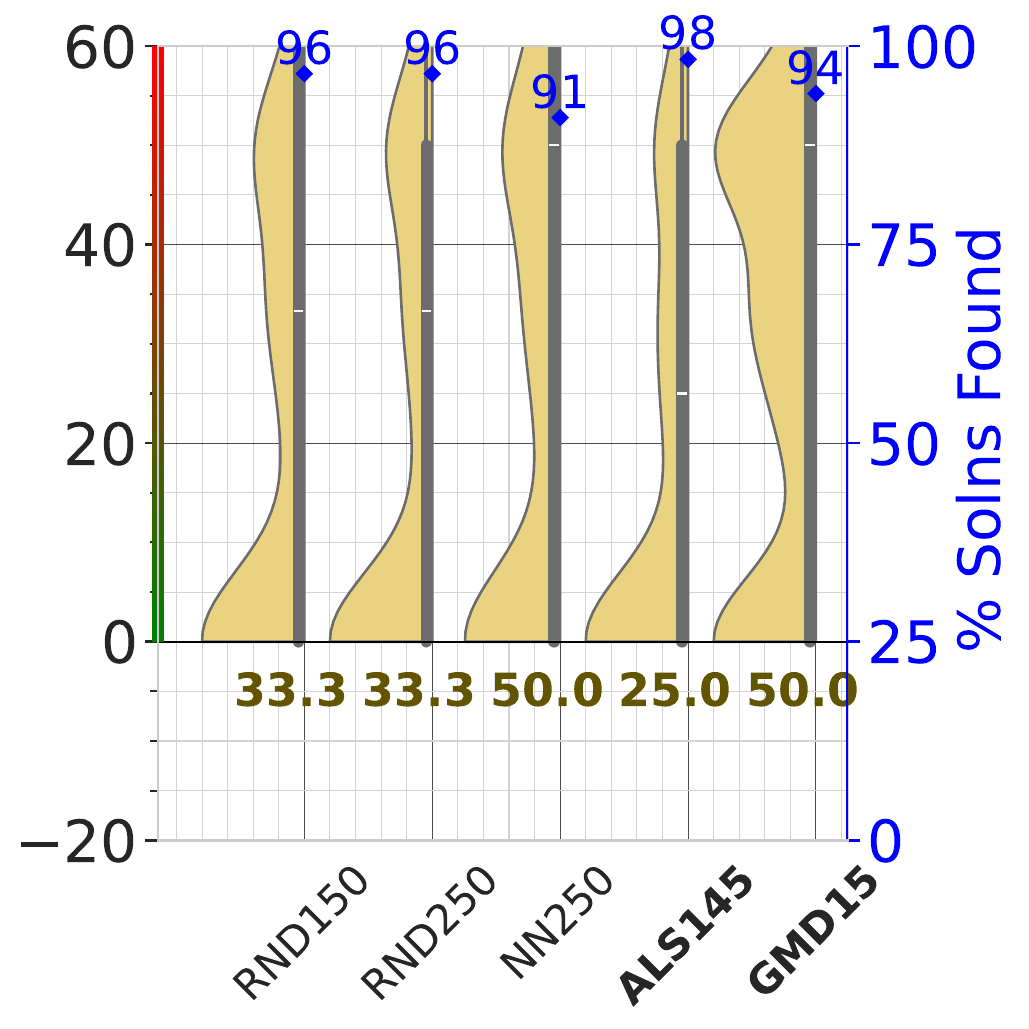}%
\label{fig:interleave_yolo_train_resnet_infer}%
}%
\vspace{-0.1in}
\caption{\% Training Throughput loss relative to optimal for all strategies and baselines on Concurrent Training and Inference workloads.}
\label{fig:interleave_Box_Plot}
\vspace{-0.15in}
\end{figure*}

We evaluate $5$ pairs out of the $25$ possible DNN training and inference combinations.
These \{\textit{train,infer}\} pairs are picked based on common tasks in popular DNN pipelines: object detection and classification \{\yolo, \resnet\},
image classification tasks \{\resnet, \mobilenet\}, \{\mobilenet, \mobilenet\}, visual question answering (VQA) or image captioning tasks which use text and vision models \{\resnet, \bert\}, video action recognition \{\mobilenet, \lstm\}.  These also cover all $5$ DNNs, either in training or in inference.
We use the same configuration for power budget, with an arrival rate range of $30$--$120$ requests per second and a latency budget of $500$ms--$2$s in steps of $100$ms. For \bert inference, the latency range is $2$--$6$s, power budget $10$--$60$W and arrival rate is $1$--$15$ RPS. This gives $\approx 6600$ problem configurations per workload-pair and $\approx 6900$ when \bert is used, resulting in a total of $\approx 33k$ configurations. 
We also evaluate our strategies on $5$ more workload pairs (not plotted) and the trends hold.
In Figure~\ref{fig:interleave_Box_Plot} we plot the training throughput loss over optimal as the violin, and \% solution found on right Y axis.

\claim{ALS145 finds solutions for $>95\%$ of the problems, offers the best training throughput among all strategies}
As seen in Figure~\ref{fig:interleave_Box_Plot}, for \{\mobilenet, \resnet\} and \{\mobilenet, \mobilenet\}, ALS145 has a median throughput loss of 0\%, matching the optimal throughput. For \{\resnet, \mobilenet\}, ALS has a median loss in throughput of $7.1\%$, which still outperforms RND150, RND250, and NN250 that are at $14.3,14.3$ and $18.2\%$. ALS has the best throughput overall
for all $5$ DNN pairs. For \{\yolo,\resnet\}, we notice that all strategies have relatively higher median loss. On examination, we see that the time available for interleaved training is low in several problems as \resnet inference and \yolo training are both computationally heavy and take significant time,
leading to low absolute training
throughput even for optimal (peak of $4$minibatches/s). This magnifies the \% throughput loss. Even here, ALS145 has the lowest loss of $25\%$.

\claim{Training throughput for GMD15 is better or comparable to RND150 in $4$ out of $5$ DNN pairs although it uses $\leq 15$ runs and finds solutions for $>93\%$ of problems}
From Figure~\ref{fig:interleave_Box_Plot}, we see that \{\mobilenet\allowbreak, \lstm\} and \{\mobilenet\allowbreak, \mobilenet\}, GMD solutions match the optimal and also outperforms RND150, but in under $15$ runs. For \{\resnet\allowbreak, \mobilenet\}, RND150 has a median throughput loss of $14.3\%$ over optimal while GMD is at $10\%$. For \{\resnet, \bert\}, both RND150 and GMD are comparable at $12.5\%$. \{\yolo, \resnet\} is the only case where GMD performs worse, but as explained above, amplifies small absolute throughput differences when expressed in \%. GMD finds solutions to $>93\%$ of configurations, finding comparable \% of solutions to RND150 in all $5$ DNNs. 

\claim{NN250 remains worse than RND250 in all $5$ DNN pairs}
From Figure~\ref{fig:interleave_Box_Plot}, RND250 matches the optimal solutions for \{\mobilenet, \lstm\} and \{\mobilenet, \mobilenet\}, while NN250 has a much higher throughput loss of $14.3\%$ and $16.7\%$. This is despite NN250 and RND250 both using the same number of 250 profiling runs. Similarly, \{\resnet, 
 \bert\} and \{\yolo\allowbreak, \resnet\}, RND250 has a median loss of $12.5$ and $33.3\%$ while NN250 is much worse at $20$ and $50\%$. This again shows that prediction errors affect the results significantly.

\subsection{Dynamic Arrival Rate}

\begin{figure*}[t]
\vspace{-0.1in}
    \centering
    \subfloat[Poisson Distribution]{%
        \includegraphics[width=0.49\textwidth]{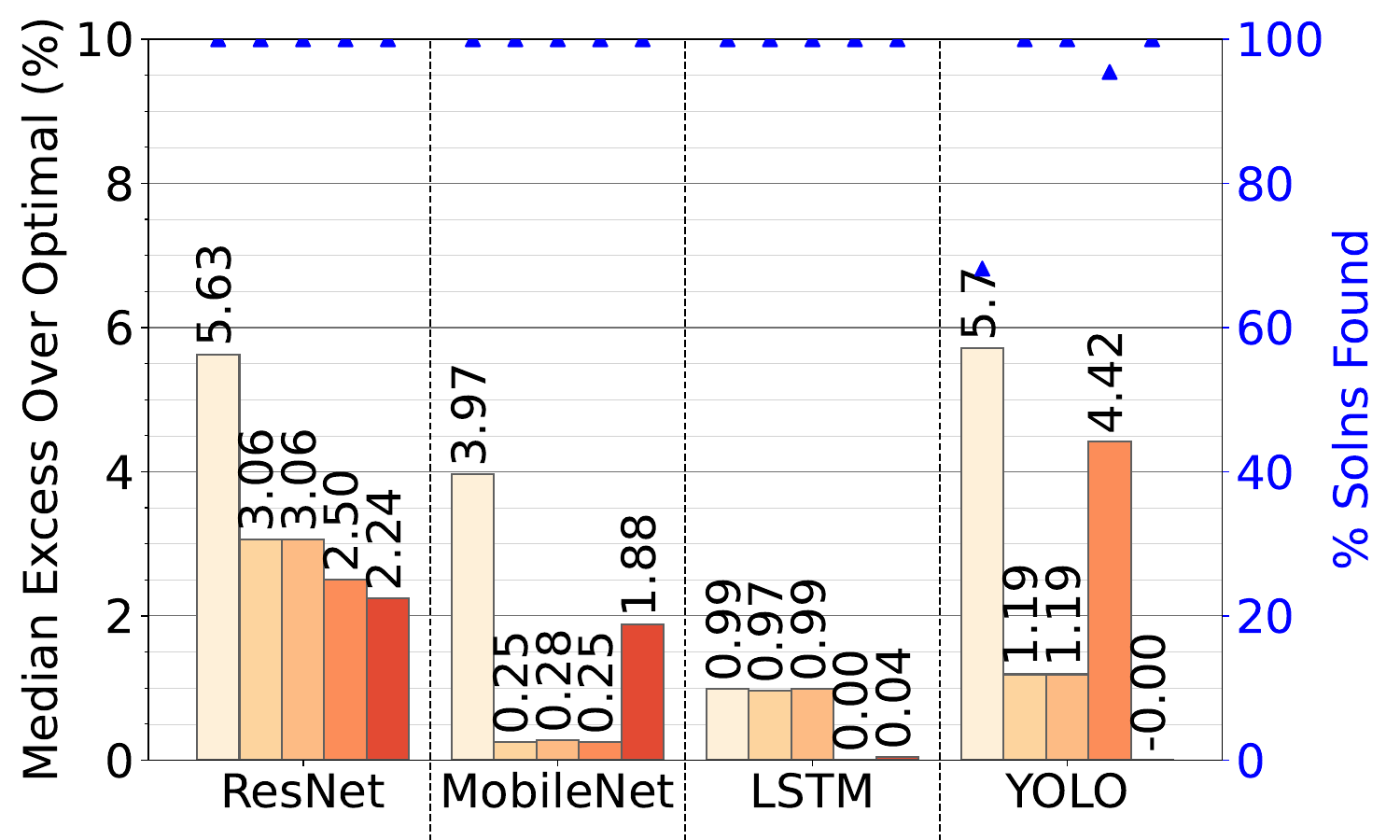}%
        \label{fig:disb_penalty}
    }\hfill
    \subfloat[Alibaba Trace]{%
        \includegraphics[width=0.49\textwidth]{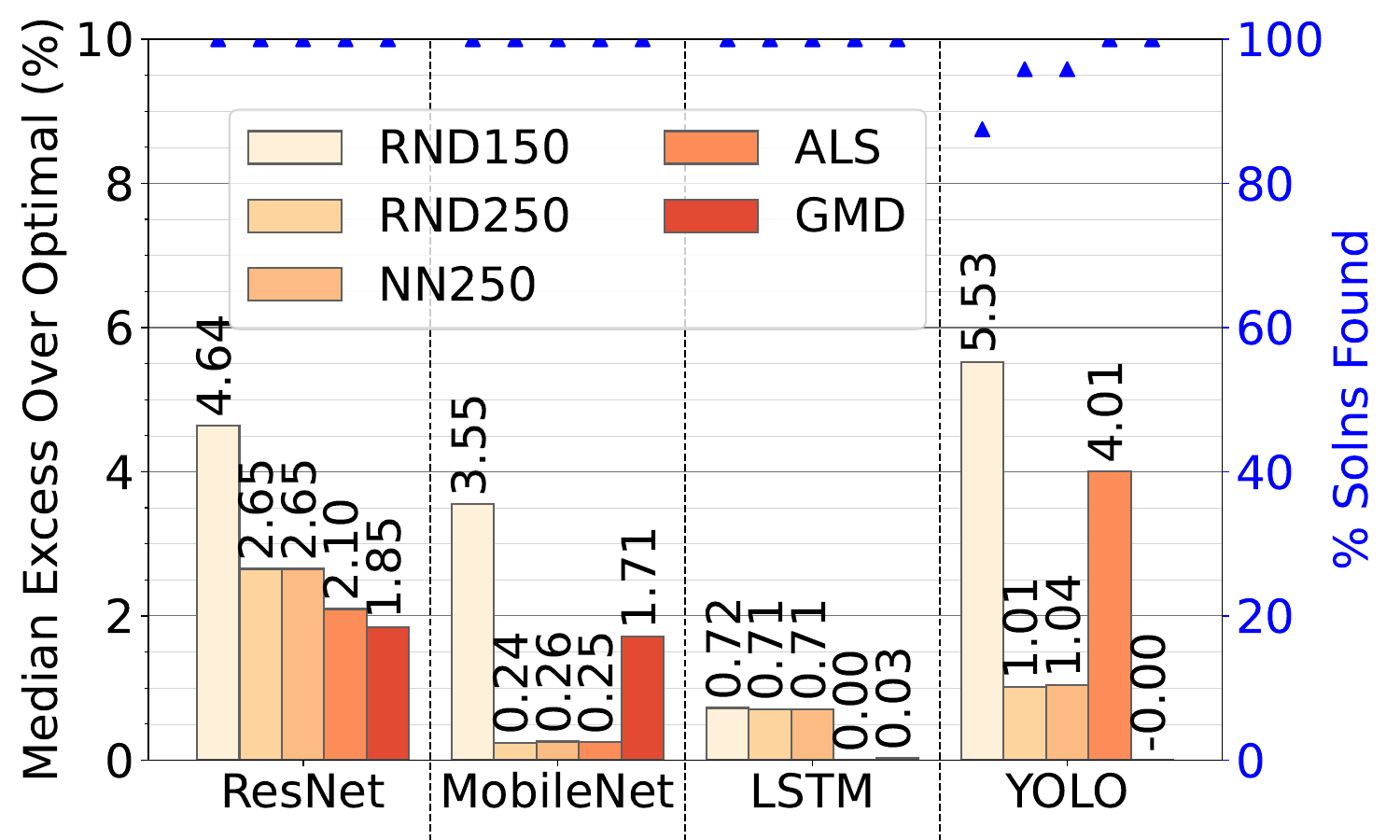}%
        \label{fig:alibaba_penalty}
    }\\
    \subfloat[Azure Trace]{%
        \includegraphics[width=0.50\textwidth]{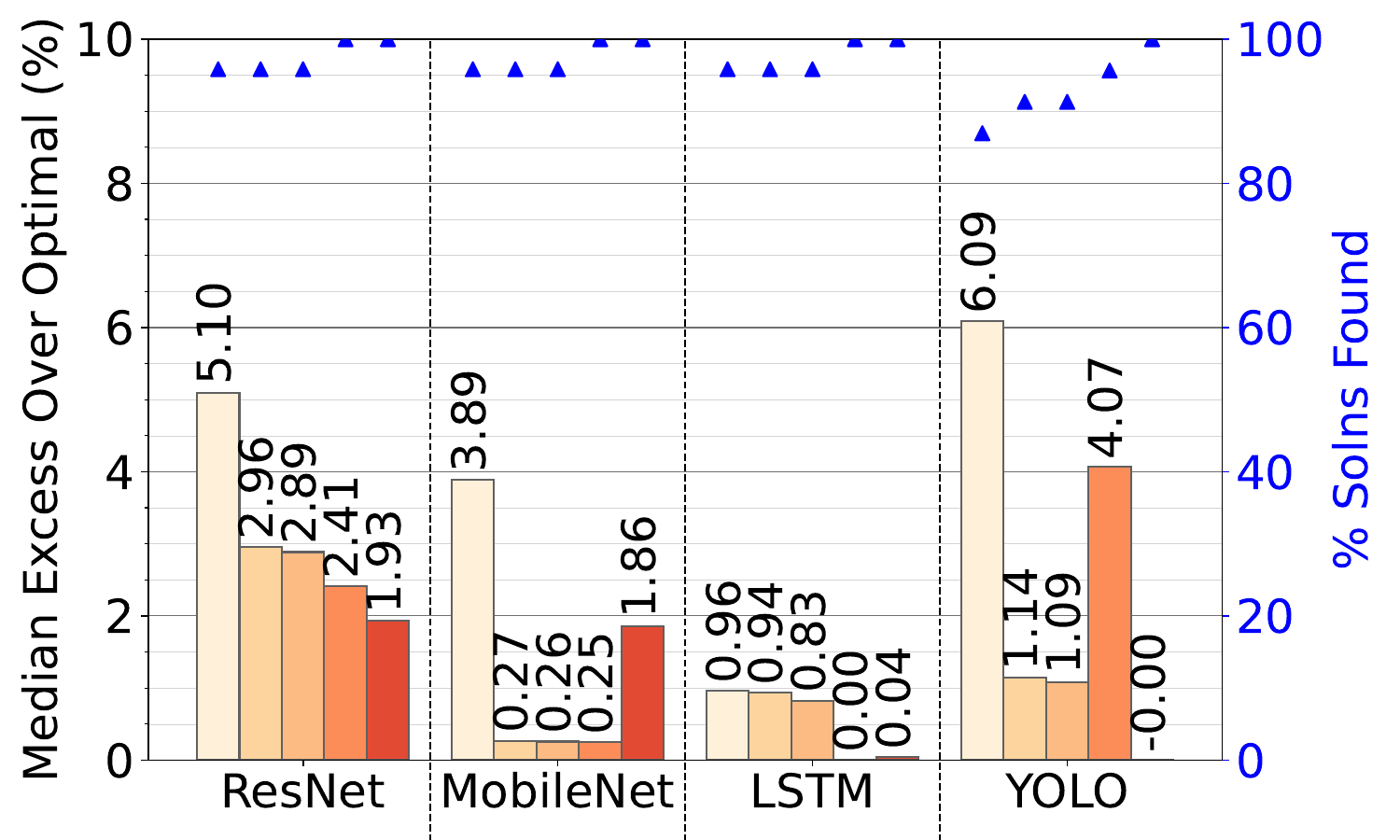}%
        \label{fig:azure_penalty}
    }
    \caption{Median excess latency over optimal inference latency (\%) for Standalone Inference with Dynamic Arrival Rates for all strategies and baselines}
    \label{fig:dynarrival}
\end{figure*}

\begin{figure}[t]
\vspace{-0.1in}
    \centering
    \subfloat[Arrival Rate (RPS) for all Dynamic inference workloads]{
        \includegraphics[width=0.5\columnwidth]{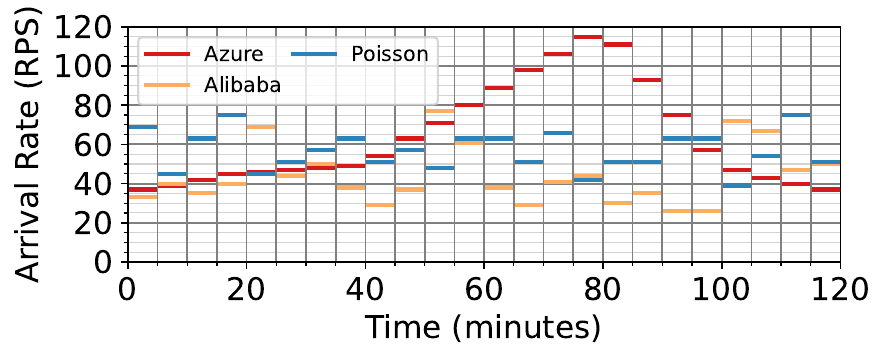}
        \label{fig:all_traces}
    }
    \subfloat[GMD and optimal latency for \resnet on the Azure trace]{
        \includegraphics[width=0.5\columnwidth]{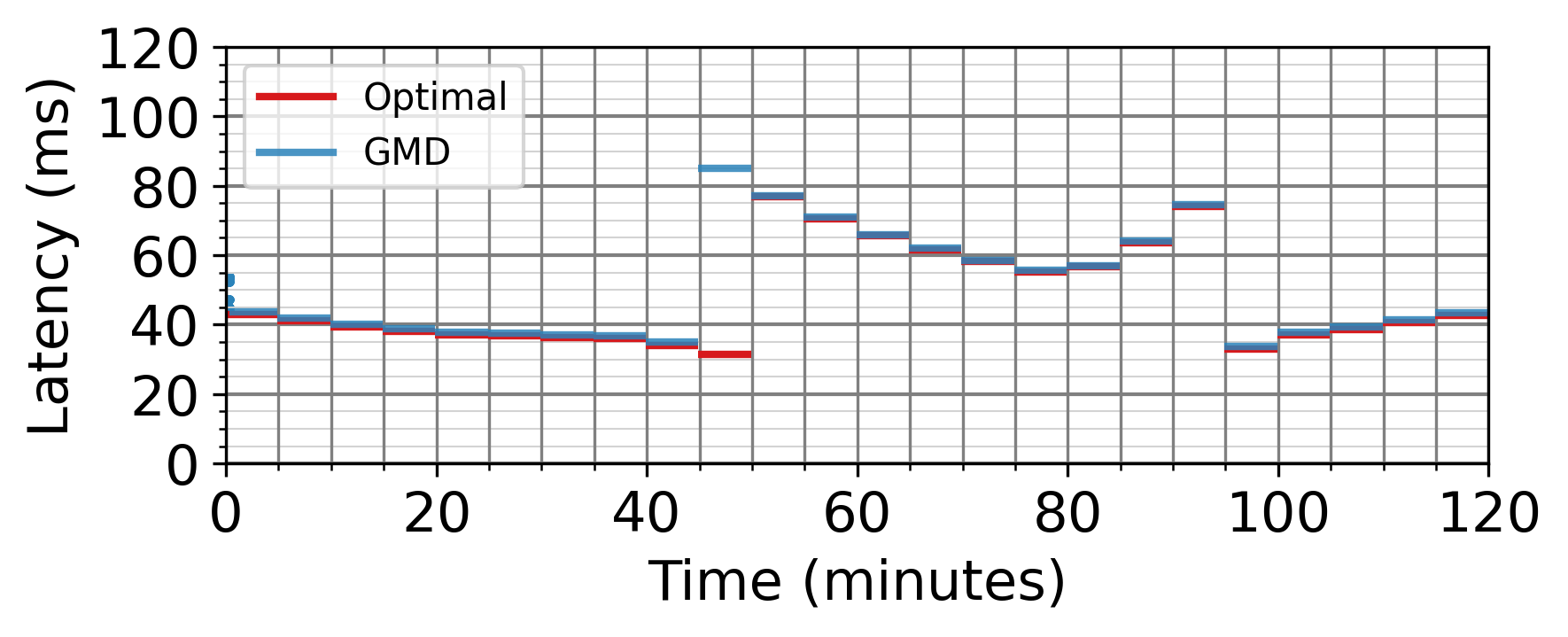}
        \label{fig:timeseries_azure}
    }
    \caption{Dynamic arrival rate inference}
    \label{fig:dyn_results}
\end{figure}

\begin{figure}[t]
\vspace{-0.1in}
    \centering
    \subfloat[\{RNet,MNet\}]{
        \includegraphics[width=0.4\columnwidth]{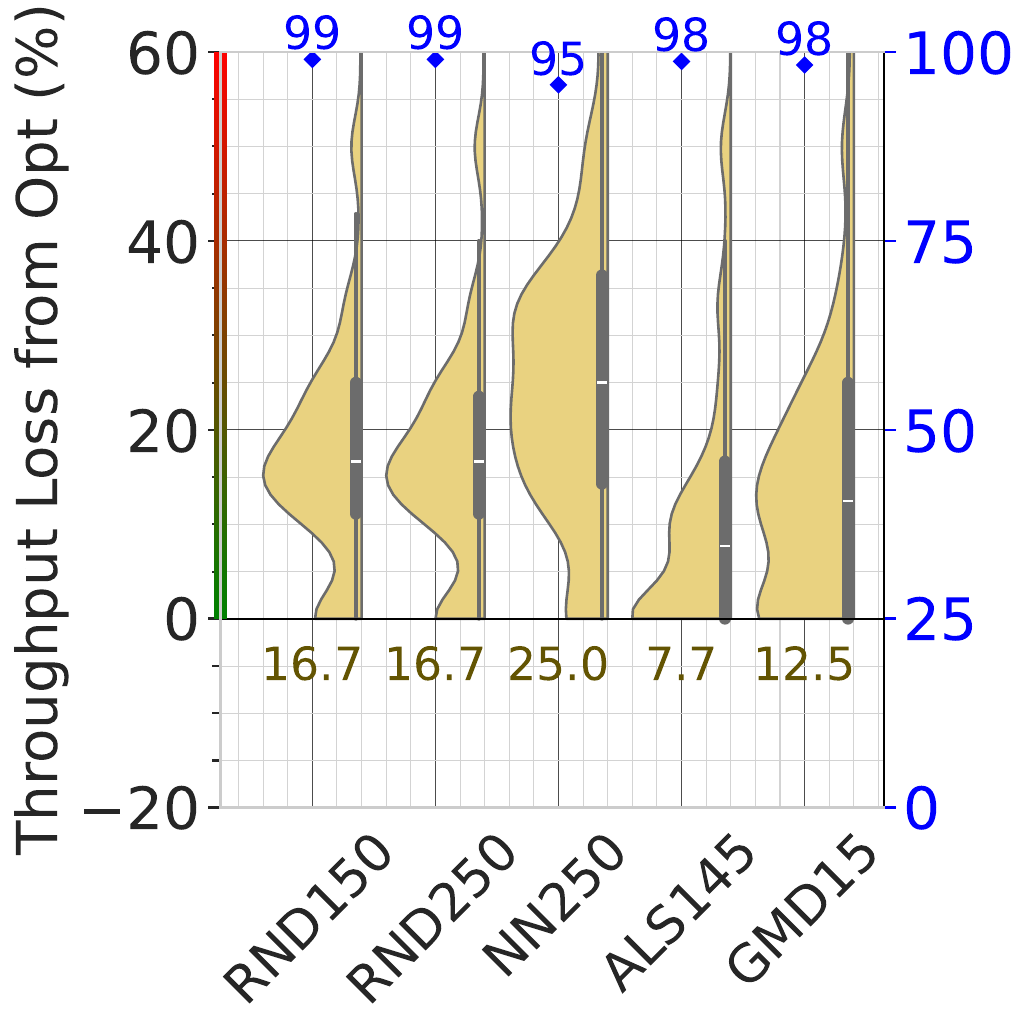}
        \label{fig:mobnet_yolo}
    }
    \subfloat[\{RNet,BERT\}]{
        \includegraphics[width=0.4\columnwidth]{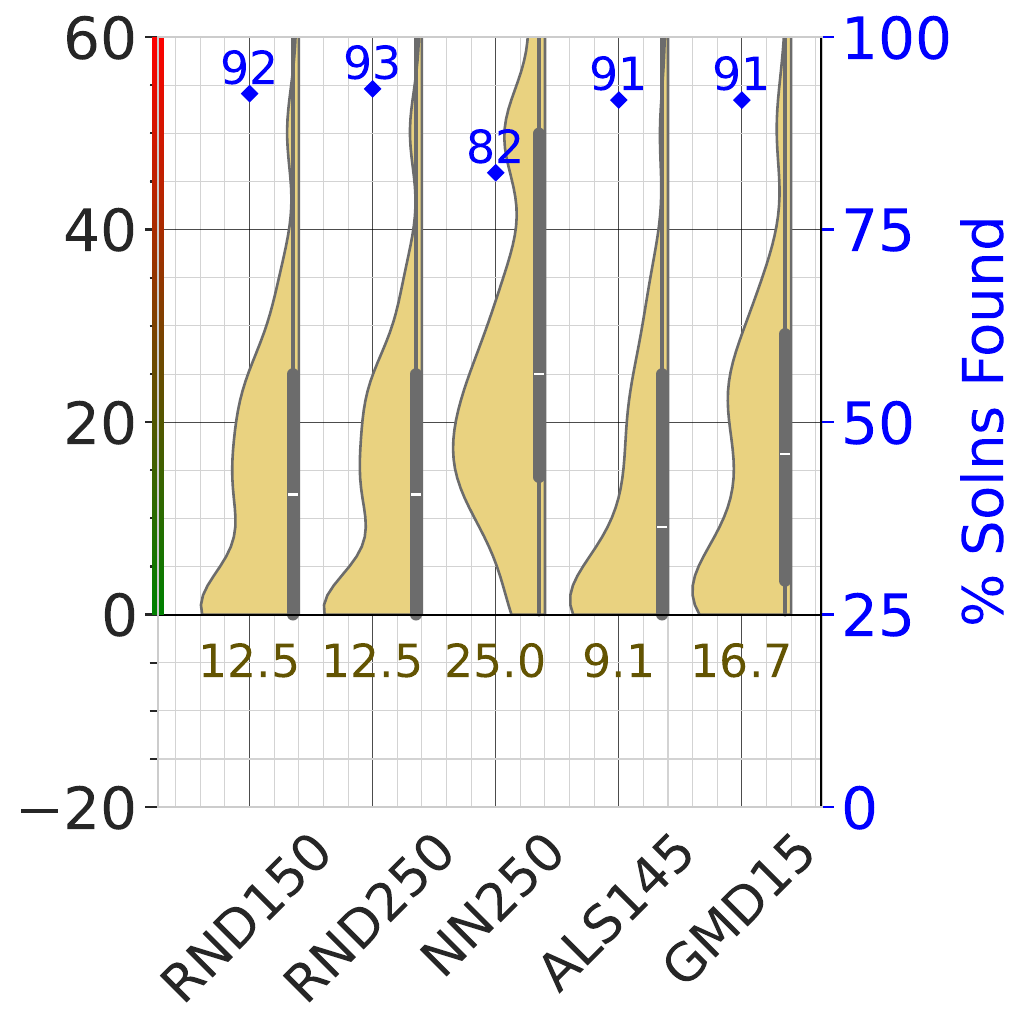}
        \label{fig:mobnet_mobnet}
    }
    \caption{\% Training Throughput loss from optimal for all strategies and baselines on Concurrent Inference workloads}
    \label{fig:concurrent_infer}
\end{figure}

We also evaluate our strategies for inference workloads with dynamic arrival rates, and compare against baselines. We evaluate this for $3$ dynamic scenarios: a \textit{Poisson distribution} of arrival rates, and two real-world traces from \textit{Alibaba's GPU cluster}~\cite{alibaba_trace} and from  \textit{Azure LLM workloads}~\cite{azure_trace}. The Poisson distribution was generated with a mean of $60$ RPS. The Alibaba and Azure traces are scaled in magnitude to be in the range of $30$-$90$ RPS that we have previously evaluated. The Poisson and Alibaba trace have a maximum arrival rate of $\approx76$ RPS
while Azure goes up to $115$ RPS, which is higher than our earlier configurations. Each trace is $2$ hours long, and the arrival rate changes every $5$ minutes (Figure~\ref{fig:all_traces}). We fix the power budget to $40$W and latency to $100$ms and replay the arrival rate as per the trace. 

\paragraph{Results}
As shown in Figure~\ref{fig:dynarrival}, we report the median excess latency over optimal (bar, left Y axis) for both strategies and the baselines on \resnet, \mobilenet, \yolo and \lstm, along with the percentage of solutions found (marker, right Y axis). We consider both the profiling and the steady state outputs of GMD while calculating the median. 

ALS has close to $0$ throughput loss for \mobilenet and \lstm, and around $2$--$2.5\%$ for \resnet. ALS outperforms all baselines on all workloads and traces except for \yolo, where its throughput loss is around $4\%$ and higher than RND250 and NN250. 
Interestingly, ALS performs equally well on Azure, despite its arrival rates exceeding the original range we evaluated for. ALS did not need further fine-tuning to adapt to these new arrival rates.

GMD outperforms RND150 in all $4$ workloads for all $3$ traces, and comparable or better throughput than RND250 and NN250 for $3$ out of $4$ workloads. It is able to find solutions $100\%$ of the time for all traces and all workloads. Figure~\ref{fig:timeseries_azure} shows a time-series plot of \resnet inference latencies using GMD (blue) and optimal (red) for the Azure arrival rate trace. The profiling time for GMD (visible at the beginning) is less than $2\%$ of the total trace time. During this profiling phase, where GMD is trying to converge to a solution, there are some dropped requests where the chosen power modes are far from optimal (spikes in the first window). After that, GMD closely follows optimal even though the latency response of optimal constantly changes. 

At $45$ min, as the arrival rate goes beyond $60$, GMD is unable to find a solution at the current batch size. It switches to a higher batch size, leading to a temporarily high difference between GMD and optimal. However, in the very next arrival rate, the optimal solution also shifts to a higher batch size, and they are both comparable. These results show that our techniques perform well on dynamic arrival rates as well.

\subsection{Concurrent Inference Workloads}

We also extend our evaluation to include two concurrent inference workloads. We evaluate over the same $6600$ configurations of latencies, arrival rates and power budgets. 

\paragraph{Results}
We evaluate two pairs of non-urgent, urgent inferences: \{RNet, BERT\} and \{RNet, MNet\}. In Figure~\ref{fig:concurrent_infer}, we present the throughput loss of ALS and GMD from optimal and compare it with the baselines. For \{RNet, MNet\}, ALS has the lowest throughput loss of $7.7\%$, followed by GMD at $12.5\%$ both of which outperform all $3$ baselines. For \{BERT, RNet\}, ALS has the least throughput loss of $9.1\%$, but GMD is $\approx4\%$ worse than RND150/RND250. NN has the highest throughput loss in both cases. These results show that our methods perform well even for multiple concurrent inferences. 

\section{Related Work}
\label{sec:related}

\paragraph{Standalone training on server and edge}

Gandiva~\cite{gandiva_osdi} uses intra-job predictability to time slice GPUs efficiently across multiple DNN training jobs and migrates jobs dynamically to better fit GPUs and improve cluster efficiency. We use time-slicing at minibatch granularities too, but between training and inference workloads. Antman~\cite{antman_osdi20} proposes a scheduling technique that uses spare resources to execute multiple jobs on a shared GPU while minimizing interference between jobs.\\
MURI~\cite{MURI_sigcomm} packs Deep Learning training jobs along multiple resource types to achieve high utilization and reduce job completion time. Zeus~\cite{zeus_nsdi23} is an optimization framework that uses an exploration--exploitation based approach to find the optimal job and GPU configurations to navigate the energy performance tradeoff of DNN training jobs. 
All these works examine server GPUs, where power is usually not a limitation, unlike in edge devices. Additionally, none of them consider CPU, GPU, and memory frequencies for optimization. 

Others~\cite{prashanthi2023sigmetrics} have characterized training of DNNs on Jetson edge devices and proposed a linear-regression based prediction model to predict training time and energy. This method is evaluated on just 10 power modes, and its errors are much higher. PowerTrain~\cite{PowerTrain} proposed a neural network based approach to predict training time and power, whose NN model we use as a baseline to solve the optimization problem. While previous works were limited to training, we consider standalone inference and concurrent training and inference too. Also, unlike prediction-based methods, the methods proposed in this work have no errors, are much closer to the optimal, and find solutions faster.

\paragraph{Concurrent training and inference on servers}
Lyra~\cite{lyra_eurosys23} proposes a scheduling mechanism to loan idle inference servers to training jobs. Ekya~\cite{ekya-stoica-nsdi} proposes a scheduler for continual learning (joint inference and retraining) on edge servers. These works too are on GPU servers and build upon GPU-sharing mechanisms such as MPS and/or MIG which Jetson edge accelerators do not support.

\paragraph{Concurrent inferences on server and edge}
GSLICE~\cite{gslice_socc20} develops a dynamic GPU partitioning scheme for multiple inference workloads on GPU servers, but it builds upon MPS that is not supported on edge GPUs. Others~\cite{multiinf_taas23} design analytical models to predict the performance of concurrent inference workloads on Jetson Nanos and TPUs. However, they do not consider power modes. MASA~\cite{masa_percom21} reduces the average response time for executing multiple DNN inference workloads with a focus on memory-constrained edge devices. Pantheon~\cite{pantheon_mobisys24} uses fine-grained preemption to support multiple DNN inferences of different priorities on an edge device. Reef~\cite{reef_concinf_osdi22} uses padding of kernels from different inference workloads to better utilize server GPUs. Unlike these works, we focus on concurrent training and inference in a power-constrained environment, and both workloads have different QoS goals and constraints.

\paragraph{Standalone inference on server and edge}
Others~\cite{matei_europar} have used a roofline model on an older generation Jetson TK1 and TX1 to characterize the performance of matrix multiplication micro-benchmarks. MIRAGE~\cite{mirage} predicts runtime and power consumption of DNN inference workloads using gradient tree boosting. ALERT~\cite{alert_atc20} selects a DNN and system resource configuration to meet latency and accuracy constraints of inference while minimizing energy on CPUs and desktop GPUs. None of these have looked at minibatch size as an additional parameter, nor do they consider CPU, GPU, and memory frequencies for optimization, or concurrent inference along with training. 

\section{Conclusion}
\label{sec:conclude}

In this article, we motivate the growing need for accelerated edge devices to execute concurrent training and inference in power-constrained deployment scenarios. We develop a managed interleaving approach within \papertitle that offers stable inferencing latency. We propose two novel optimization strategies GMD and ALS that satisfy the diverse QoS goals of meeting inference latency and maximizing training throughput while staying within a power budget for field deployments. These are profiling-based strategies that do not suffer from prediction errors unlike the NN baseline. They outperform both random and NN baselines while requiring limited profiling.

As future work, we plan to look into concurrency levels higher than two and develop a scheduler that meets power constraints. We also plan to investigate strategies that use Reinforcement Learning and Multi-Armed Bandits for optimization.

\bibliographystyle{IEEEtran}
\bibliography{arxiv}
\balance


\section{Appendix}

\begin{algorithm}[h]
\caption{Gradient-descent based Multi-Dimensional search (GMD) for Training Workload}
\label{algo:gmd}
\footnotesize
    \begin{algorithmic}[1]
    \Procedure{GradientMultiDimSearch\_Train} {$\mu_{tr}, \widehat{p}$}
    \State $best\_t_{tr} \gets \infty$, $best\_p \gets 0$, $best\_pm \gets \phi$ \label{algo_gmd:init:start}
    \State $tries \gets 0$, $max\_tries \gets 10$
    \State $dims = \{ \text{CORES}, \text{CPUF}, \text{GPUF}, \text{MEMF} \}$
    \State $cores = \langle core_{low}, core_{high} \rangle$, $cpuf = \langle cpuf_{low}, cpuf_{high} \rangle$, $gpuf = \langle gpuf_{low}, gpuf_{high} \rangle$, $memf = \langle memf_{low}, memf_{high} \rangle$

    \State $space = cores \times cpuf \times gpuf \times memf$

    \State $pm^\text{mid} \gets  \langle \Call{mid}{cores}, \Call{mid}{cpuf}, \Call{mid}{gpuf}, \Call{mid}{memf} \rangle$ \label{algo_gmd:init:end}
    
    \State $p_\text{tr}^\text{mid}, t_\text{tr}^\text{mid} \gets \Call{Profile}{pm^\text{mid}}$ \label{algo_gmd:mid:start}
    \If{$p_\text{tr}^\text{mid} > \hat{p}$}
        \State $pm[\text{CORES}] \gets core_\text{low}, \Call{mid}{cpuf}, \Call{mid}{gpuf}, \Call{mid}{memf}$
        \State $pm[\text{CPUF}] \gets \Call{mid}{cores}, cpuf_\text{low}, \Call{mid}{gpuf}, \Call{mid}{memf}$
        \State $pm[\text{GPUF}] \gets \Call{mid}{cores}, \Call{mid}{cpuf}, gpuf_\text{low}, \Call{mid}{memf}$
        \State $pm[\text{MEMF}] \gets \Call{mid}{cores}, \Call{mid}{cpuf}, \Call{mid}{gpuf}, memf_\text{low}$
        \State $prune\_dir \gets \text{high}$

    \Else
        \State $pm[\text{CORES}] \gets core_\text{high}, \Call{mid}{cpuf}, \Call{mid}{gpuf}, \Call{mid}{memf}$
        \State $pm[\text{CPUF}] \gets \Call{mid}{cores}, cpuf_\text{high}, \Call{mid}{gpuf}, \Call{mid}{memf}$
        \State $pm[\text{GPUF}] \gets \Call{mid}{cores}, \Call{mid}{cpuf}, gpuf_\text{high}, \Call{mid}{memf}$
        \State $pm[\text{MEMF}] \gets \Call{mid}{cores}, \Call{mid}{cpuf}, \Call{mid}{gpuf}, memf_\text{high}$
        \State $prune\_dir \gets \text{low}$
    \EndIf \label{algo_gmd:mid:end}

    \For {$dim \in dims$} \label{algo_gmd:profile:start}
        \State $p_{tr}[dim], t_{tr}[dim] \gets \Call{Profile}{pm[dim]}$
        \State $m^\text{pow}_\text{dim}, m^\text{time}_\text{dim} \gets \Call{CalcSlope}{pm[dim],p_{tr}[dim], t_{tr}[dim], pm^{mid}, \-p_{tr}^{mid}, t_{tr}^{mid}}$
        \State $\rho_\text{dim} \gets \frac{m^\text{time}_\text{dim}}{m^\text{pow}_\text{dim}}$
    \EndFor \label{algo_gmd:profile:end}

    \State $dim \gets \Call{argmax}{\rho}$ \label{algo_gmd:prune:start}

    \State $space, dim\_vals \gets \Call{Prune}{dim, prune\_dir}$
    
    \While{$(tries < max\_tries) \text{ OR } (|space| > 1)$}

        \State $pm^{(1)} \gets \Call{MID}{dim\_vals}$, rest unchanged
        \State $pm^{(2)} \gets pm[dim]$ 
        \State $p_{tr}^{(2)}, t_{tr}^{(2)} = p_{tr}[dim], t_{tr}^{mid}[dim]$

        \State $p_{tr}, t_{tr} \gets \Call{Profile}{pm^{(1)}}$
        \If{$p_{tr} > \hat{p}$}
            \State $prune\_dir \gets high$
            \State $space, dim\_vals \gets \Call{Prune}{dim, prune\_dir}$
        \Else
            \If{$(t_{tr} < best\_t_{tr})$}  
                \State $best\_{pm} \gets pm^{(1)}, best\_t_{tr} \gets t_{tr}, best\_p \gets p_{tr}$
            \EndIf
            \State $prune\_dir \gets low$
            \State $space, dim\_vals \gets \Call{Prune}{dim, prune\_dir}$
        \EndIf
        \State $m^\text{pow}_\text{dim}, m^\text{time}_\text{dim} \gets \Call{CalcSlope}{pm^{(1)}, p_{tr}[dim], t_{tr}[dim], pm^{(2)}, p_{tr}^{(2)}, t_{tr}^{(2)}}$
        \State $dim \gets \Call{argmax}{\rho}$
    \State $tries \gets tries + 1$
    \EndWhile \label{algo_gmd:prune:end}

     \State \Return $best\_pm, best\_t_{tr}, best\_p$
\EndProcedure
\end{algorithmic}
\end{algorithm}


\begin{algorithm}[h]
\small
\caption{Active Learning-based Sampling (ALS) for Training Workload}
\label{algo:als}
\begin{algorithmic}[1]

\Procedure{ALS\_Train}{$\mu_{tr}$}
    \State $max\_rnds \gets 8$, $num\_pms\_init \gets 10$, $num\_pms\_rnd \gets 5$
    
    \State $\mathcal{M} \gets \Call{RandWeights}{ }$ \label{alg2:train:start}
    \State $PM^{(\delta)}_{\text{train}} \gets \Call{PickRand}{PM_{\text{all}}, num\_pms\_init}$ 
    \State $PM_{\text{train}} \gets PM^{(\delta)}_{\text{train}}$
    \State $PM_{\text{test}} \gets PM_{\text{all}} - PM_{\text{train}}$

    \For{$r = 1$ \textbf{to} $max\_rnds + 1$}
        
        \For{$pm_{\text{train}} \in PM^{(\delta)}_{\text{train}}$}
            \State $t_{\text{obs}}[pm_{\text{train}}], p_{\text{obs}}[pm_{\text{train}}] \gets \Call{Profile}{pm_{\text{train}}}$

        \EndFor
        
        \State $\mathcal{M} \gets \Call{NNTrain}{PM^{(\delta)}_{\text{train}}, t_{\text{obs}}, p_{\text{obs}}, \mathcal{M}}$ \label{alg2:train:end}

        \For{$pm_{\text{test}} \in PM_{\text{test}}$} \label{alg2:pred:start}
            \State $t_{\text{pred}}[pm_{\text{test}}], p_{\text{pred}}[pm_{\text{test}}] \gets \mathcal{M}(pm_{\text{test}})$
        \EndFor

        \State $PM_{\text{pareto}} \gets \Call{ParetoEfficient}{PM_{\text{test}}, t_{\text{pred}}, p_{\text{pred}}}$ \label{alg2:pred:end}

        \For{$pm_{\text{pareto}} \in PM_{\text{pareto}}$} \label{alg2:greedy:start}
            \For{$pm_{\text{train}} \in PM_{\text{train}}$}
                \State $dist[pm_{\text{pareto}}][pm_{\text{train}}] \gets |p_{\text{pred}}[pm_{\text{pareto}}] - p_{\text{obs}}[pm_{\text{train}}]|$
            \EndFor
            \State $min\_dist[pm_{\text{pareto}}] \gets \min(dist[pm_{\text{pareto}}])$
        \EndFor

        \State $PM^{(\delta)}_{\text{train}} \gets \Call{PickMaxK}{PM_{\text{pareto}}, min\_dist, num\_pms\_rnd}$ \label{alg2:greedy:end}
        \State $PM_{\text{train}} \gets PM_{\text{train}} \cup PM^{(\delta)}_{\text{train}}$
        \State $PM_{\text{test}} \gets PM_{\text{all}} - PM_{\text{train}}$
    \EndFor

    \State \Return $PM_{\text{train}}$
\EndProcedure

\end{algorithmic}
\end{algorithm}

\end{document}